\documentclass[aps,prd,superscriptaddress,floatfix,twocolumn,nofootinbib,preprintnumbers]{revtex4}

\usepackage{amsmath,amssymb,amsfonts,dcolumn,slashed,dsfont,multirow}
\usepackage{graphicx}
\usepackage{psfrag,nicefrac,braket}
\usepackage{bold-extra}

\renewcommand{\Re}{\text{Re}\,}
\newcommand{\mpi}{M_{\pi}}

\newcommand{\meta}{M_\eta}
\newcommand{\Fpi}{F_\pi}
\newcommand{\gA}{g_A}
\newcommand{\beq}{\begin{equation}}
\newcommand{\eeq}{\end{equation}}

\newcommand{\mN}{m_N}
\newcommand{\mA}{m_A}

\newcommand{\qq}{\mathbf{q}}
\newcommand{\pp}{\mathbf{p}}
\newcommand{\PP}{\mathbf{P}}
\newcommand{\kk}{\mathbf{k}}
\newcommand{\JJ}{\mathbf{J}}
\newcommand{\llvec}{\mathbf{l}}

\newcommand{\N}{\mathcal{N}}
\newcommand{\M}{\mathcal{M}}
\newcommand{\F}{\mathcal{F}}
\newcommand{\spin}{\mathbf{S}}
\newcommand{\unity}{\mathds{1}}
\newcommand{\sig}{\boldsymbol{\sigma}}

\newcommand{\bfnabla}{\boldsymbol{\nabla}}
\newcommand{\rr}{\boldsymbol{r}}
\newcommand{\diff}{\text{d}}
\newcommand{\eps}{\epsilon}
\newcommand{\Order}{\mathcal{O}}

\newcommand{\Lagr}{\mathcal L}
\newcommand{\Tr}{\text{Tr}}

\providecommand{\MeV}{\,\text{MeV}}
\providecommand{\GeV}{\,\text{GeV}}
\providecommand{\fm}{\,\text{fm}}

\allowdisplaybreaks[1]

\begin{document}

\preprint{INT-PUB-20-026}

\title{Coherent elastic neutrino--nucleus scattering: EFT analysis and nuclear responses}

\author{Martin Hoferichter}
\email[E-mail:~]{hoferichter@itp.unibe.ch}
\affiliation{Albert Einstein Center for Fundamental Physics, Institute for Theoretical Physics, University of Bern, Sidlerstrasse 5, 3012 Bern, Switzerland}
\affiliation{Institute for Nuclear Theory, University of Washington, Seattle, WA 98195-1550, USA}

\author{Javier Men\'endez}
\email[E-mail:~]{menendez@fqa.ub.edu}
\affiliation{Department of Quantum Physics and Astrophysics
and Institute of Cosmos Sciences,
University of Barcelona, Spain}
\affiliation{Center for Nuclear Study, The University of Tokyo, 113-0033 Tokyo, Japan}

\author{Achim Schwenk}
\email[E-mail:~]{schwenk@physik.tu-darmstadt.de}
\affiliation{Institut f\"ur Kernphysik, 
Technische Universit\"at Darmstadt, 
64289 Darmstadt, Germany}
\affiliation{ExtreMe Matter Institute EMMI, 
GSI Helmholtzzentrum f\"ur Schwerionenforschung GmbH, 
64291 Darmstadt, Germany}
\affiliation{Max-Planck-Institut f\"ur Kernphysik, Saupfercheckweg 1, 
69117 Heidelberg, Germany}

\begin{abstract}
The cross section for coherent elastic neutrino--nucleus scattering (CE$\nu$NS) depends on the response of the target nucleus to the external current, 
in the Standard Model (SM) mediated by the exchange of a $Z$ boson. This is typically subsumed into an object called the weak form factor of the nucleus. 
Here, we provide results for this form factor calculated using the large-scale nuclear shell model
for a wide range of nuclei of relevance for current CE$\nu$NS experiments, including cesium, iodine, argon, fluorine, sodium, germanium, and xenon. 
In addition, we provide the responses needed to capture the axial-vector part of the cross section, which does not scale coherently with the number of neutrons, 
but may become relevant for the SM prediction of CE$\nu$NS on target nuclei with nonzero spin.
We then generalize the formalism allowing for contributions beyond the SM. In particular, 
we stress that in this case, even for vector and axial-vector operators, the standard weak form factor does not apply anymore,   
but needs to be replaced by the appropriate combination of the underlying nuclear structure factors. 
We provide the corresponding expressions for vector, axial-vector, but also (pseudo-)scalar, tensor, and dipole effective operators, including two-body-current effects as predicted from chiral effective field theory.
Finally, we update the spin-dependent structure factors for dark matter scattering off nuclei according to our improved treatment of the axial-vector responses.  
\end{abstract}

\maketitle

\section{Introduction}
\label{sec:intro}

CE$\nu$NS, suggested as a probe of the weak current as early as 1974~\cite{Freedman:1973yd}, 
was finally observed by the COHERENT collaboration in 2017~\cite{Akimov:2017ade}. After the initial 
detection in CsI, also the scattering off argon has just been observed~\cite{Akimov:2019rhz,Akimov:2020pdx}.
With future advances in COHERENT and other experiments~\cite{Belov:2015ufh,Kerman:2016jqp,Agnolet:2016zir,Akimov:2017hee,Hakenmuller:2019ecb,Angloher:2019flc,Aguilar-Arevalo:2019jlr,Juillard:2019njs,Baxter:2019mcx},
the CE$\nu$NS process will soon develop into another sensitive low-energy probe of physics beyond the Standard Model (BSM)~\cite{Barranco:2005yy}.

A crucial input in interpreting the measured cross section is the response of the nucleus. If BSM constraints are to be extracted, the nuclear structure has to be provided from elsewhere. In fact, since the weak charge of the proton is small, the SM CE$\nu$NS process mainly probes the nuclear neutron distribution, which is significantly less constrained experimentally than the electromagnetic charge distribution.
Apart from CE$\nu$NS, the only direct information of the neutron distribution comes from parity-violating electron scattering (PVES) off lead~\cite{Abrahamyan:2012gp,Horowitz:2012tj}. 
Accordingly, assuming the absence of a significant BSM signal, the measured CE$\nu$NS cross section could be used to constrain the neutron distribution instead~\cite{Cadeddu:2017etk,Huang:2019ene,Khan:2019cvi,Cadeddu:2019eta,Coloma:2020nhf}.
Currently, the nuclear input used in the interpretation of CE$\nu$NS experiments is mainly derived from relativistic mean-field methods (RMF)~\cite{Horowitz:2003cz,Yang:2019pbx},
even though results based on nonrelativistic energy-density functionals are also available~\cite{Patton:2012jr,Co:2020gwl,VanDessel:2020}.
For argon, there is a recent first-principles calculation based on coupled-cluster theory~\cite{Payne:2019wvy}.

Here, we provide nuclear structure results for CE$\nu$NS, extending large-scale nuclear shell model calculations developed in the context 
of nuclear structure factors in direct-detection searches for dark matter~\cite{Menendez:2012tm,Klos:2013rwa,Baudis:2013bba,Vietze:2014vsa,Hoferichter:2015ipa,Hoferichter:2016nvd,Hoferichter:2017olk,Fieguth:2018vob,Hoferichter:2018acd}. 
First, the level of agreement between the shell-model, the RMF, and coupled-cluster results suggests that the form factor uncertainties are not as severe as claimed in Ref.~\cite{AristizabalSierra:2019zmy}, 
but in addition the shell-model approach also allows us to address the spin-dependent (SD) responses, which are similar, but somewhat different to the ones in SD dark matter searches.

To this end, we first derive the decomposition of the cross section into Wilson coefficients of effective operators, hadronic matrix elements, and nuclear structure factors. 
In the SM the effective operators just parameterize the $Z$-boson exchange, but this approach can be conveniently extended to include BSM effects. The hadronic matrix elements 
determine the hadronization of these operators at the single-nucleon level, and finally the nuclear structure factors take into account the many-body nuclear matrix element of these single-nucleon currents.
As a first step, we demonstrate how the standard weak form factor emerges when combining all these ingredients into a single object. However, this analysis shows that even 
for the coherent part of the nuclear response four different underlying structure factors contribute to the cross section.
Therefore, in principle the weak form factor needs to be modified as well when allowing 
for BSM effects in the Wilson coefficients, since their contribution does not factorize.

While for the dominant vector operators corrections beyond the single-nucleon currents are small~\cite{Bacca:2014tla,NevoDinur:2018hdo}, since 
the magnetic-moment form factors happen to be kinematically suppressed, 
this is no longer true for the axial-vector~\cite{Klos:2013rwa} or for scalar currents, see~\cite{Prezeau:2003sv,Cirigliano:2012pq,Cirigliano:2013zta,Hoferichter:2015ipa,Hoferichter:2016nvd,Hoferichter:2017olk,Hoferichter:2018acd,Korber:2017ery,Andreoli:2018etf,Aprile:2018cxk} for the analogous effects in the case of dark matter scattering off nuclei. As long as the SM dominates, such effects will only become relevant for CE$\nu$NS once experiments become sensitive to SD responses. Otherwise, 
mainly the limits on scalar operators would be affected, but in CE$\nu$NS such contributions are suppressed due to the need for right-handed neutrinos and lack of interference with the SM.   

The paper is organized as follows: in Sec.~\ref{sec:formalism} we first review the necessary formalism, regarding effective operators, hadronic matrix elements, and nuclear responses, with some details of the multipole expansion and nuclear-structure calculations summarized in the appendices. In Sec.~\ref{sec:SM} we first introduce the charge and weak form factors in the context of electron scattering, before discussing the application to CE$\nu$NS. In particular, we present an improved treatment of the axial-vector responses both for CE$\nu$NS and dark matter. In Sec.~\ref{sec:BSM} we discuss how the nuclear responses need to be adapted when considering SM extensions, before concluding with a summary in Sec.~\ref{sec:summary}.

\section{Formalism}
\label{sec:formalism}

\subsection{Effective Lagrangians}

As a first step, we review the operator basis for CE$\nu$NS~\cite{Hoferichter:2018acd,Altmannshofer:2018xyo}\footnote{This definition strictly applies to Dirac neutrinos, while in the Majorana case a symmetry factor of $2$ would arise in the amplitudes. To avoid this complication, an additional factor of $1/2$ is implied in the definition of the effective operators for Majorana neutrinos, in which case also the diagonal vector and tensor currents vanish.}
\begin{align}
 \Lagr^{(5)}&=C_F \bar\nu \sigma^{\mu\nu} P_L\nu F_{\mu\nu},\notag\\
 \Lagr^{(6)}&=\sum_{q}\Big(C_q^V\bar\nu\gamma^\mu P_L\nu \,\bar q\gamma_\mu q
 +C_q^A\bar\nu\gamma^\mu P_L\nu \,\bar q\gamma_\mu\gamma_5 q\notag\\
 &+C_q^T\bar\nu \sigma^{\mu\nu}P_L\nu\,\bar q\sigma_{\mu\nu}q\Big),\notag\\
 \Lagr^{(7)}&=\sum_q\Big(C_q^S+\frac{8\pi}{9}C_g'^S\Big)\bar\nu P_L\nu\, m_q\bar q q\notag\\
 &+\sum_q C_q^P\bar\nu P_L\nu\, m_q\bar qi\gamma_5 q -\frac{8\pi}{9}C_g'^S\bar\nu P_L\nu\,\theta^\mu_\mu,
 \label{operators_basis}
\end{align}
where we adopted the following conventions: neutrino indices are suppressed throughout, indicating that oscillation effects are usually negligible at the scale of CE$\nu$NS experiments, so that 
incoming and outgoing flavors are understood to be identical. In comparison 
to the case of a dark-matter spin-$1/2$ particle~\cite{Hoferichter:2018acd} the number of operators is reduced by a factor of $2$ when assuming that the neutrino is left-handed. This is implemented in Eq.~\eqref{operators_basis} in terms of projectors $P_L=(1-\gamma_5)/2$, and given that observing chirality-violating effects would require right-handed neutrino beams (suppressed by tiny neutrino masses), we will not consider the opposite chirality in the following.

With these conventions the set of operators is then similar to the dark-matter case: at dimension-$5$ level there is a single (dipole) operator involving the photon field strength tensor $F_{\mu\nu}$. At dimension $6$ we have the vector and axial-vector operators already present in the SM, as well as the tensor operator. Introducing quark masses for renormalization-group invariance, the scalar and pseudoscalar operators would be counted as dimension-$7$. The operator involving the QCD trace anomaly $\theta^\mu_\mu$ would also be of dimension 7.
We have already rewritten the gluon term $G^a_{\mu\nu}G^{\mu\nu}_a$ in terms of this operator  (we will not consider the operators involving the dual field strength tensor $\tilde G_{\mu\nu}^a$ or the photon field strength). In particular, we already integrated out the heavy quarks~\cite{Shifman:1978zn} and absorbed their effect into
\beq
C_g'^S=C_g^S-\frac{1}{12\pi}\sum_{Q=c,b,t}C_Q^{S},
\eeq
where $C_g^S$ is the original coefficient of the $\bar\nu P_L\nu\,\alpha_s G^a_{\mu\nu}G^{\mu\nu}_a$ gluon operator and we used the relation
\beq
\theta^\mu_\mu=\sum_q m_q\bar q q-\frac{9}{8\pi}\alpha_s G^a_{\mu\nu}G^{\mu\nu}_a+\Order\big(\alpha_s^2\big)
\eeq
in the transition. Elsewhere, the sum over $q$ in principle refers to all quark species, but in practice we will restrict the analysis to the light quarks $q=u,d,s$. Finally, there are operators with derivatives acting on the neutrino fields (in analogy to the spin-$2$ operator for dark matter), but we will concentrate on the more frequently considered operators in Eq.~\eqref{operators_basis}. We note that the dimensional counting is not unambiguous regarding the quark mass $m_q$, e.g., sometimes the tensor operator is introduced at dimension-$7$ by adding a factor $m_q$ in this operator as well~\cite{Altmannshofer:2018xyo} (the notation in Eq.~\eqref{operators_basis} follows Ref.~\cite{Goodman:2010ku}). Finally, we stress that the chirality-flipping operators, with scalar and tensor operators on the neutrino bilinear, require the presence of (final-state) right-handed neutrinos. In SMEFT~\cite{Grzadkowski:2010es} such operators are suppressed beyond dimension-$6$ level. In addition, the dipole operator leads to a new long-range interaction, and therefore $C_F$ is strongly constrained by astrophysical observations~\cite{Mohapatra:1998rq,Giunti:2014ixa}. However, such operators have been suggested as a potential BSM explanation of the excess of electronic recoil events observed by XENON1T~\cite{Aprile:2020}.

In the SM all Wilson coefficients except for $C_q^V$ and $C_q^A$ vanish, with $Z$ exchange leading to the matching relations
\begin{align}
\label{Wilson_SM}
 C_u^V=-\frac{G_F}{\sqrt{2}}\bigg(1-\frac{8}{3}\sin^2\theta_W\bigg),\notag\\
 C_d^V=C_s^V=\frac{G_F}{\sqrt{2}}\bigg(1-\frac{4}{3}\sin^2\theta_W\bigg),\notag\\
 C_u^A=-C_d^A=-C_s^A=\frac{G_F}{\sqrt{2}},
\end{align}
with Fermi constant $G_F=1.1663787(6)\times 10^{-5}\GeV^{-2}$~\cite{Tishchenko:2012ie,Tanabashi:2018oca}. In the notation of Refs.~\cite{Barranco:2005yy,Akimov:2017ade,Akimov:2020pdx}, the deviations from these SM values are often expressed as
\begin{align}
\label{BSM_epsV}
 C_q^V-C_q^V\big|_\text{SM}&=-\sqrt{2}G_F\eps^{qV}_{ee}=-\sqrt{2}G_F\big(\eps^{qL}_{ee}+\eps^{qR}_{ee}\big),\notag\\
C_q^A-C_q^A\big|_\text{SM}&=\sqrt{2}G_F\eps^{qA}_{ee}=\sqrt{2}G_F\big(\eps^{qL}_{ee}-\eps^{qR}_{ee}\big),
\end{align}
where the sign of $\eps^{qA}_{ee}$ has been chosen in accordance with Ref.~\cite{Barranco:2005yy}. Finally, we can define dimensionless Wilson coefficients $\tilde C_i=C_i/\Lambda^n$, where $\Lambda$ either corresponds to the respective 
BSM scale, or, in the SM, to the Higgs vev $\Lambda=(\sqrt{2} G_F)^{-1/2}=v=246\GeV$.

\subsection{Dimension-$\boldsymbol{5}$ matrix elements}

Having defined the operator basis~\eqref{operators_basis}, the second step concerns the nonperturbative input required to define amplitudes at the hadronic level. We will largely follow the conventions of Refs.~\cite{Hoferichter:2015ipa,Hoferichter:2018acd}, but for completeness review here the respective hadronic matrix elements. 

For the dimension-$5$ operator only the electromagnetic form factors of the nucleon are required, without the need for a flavor decomposition. With $N=\{p,n\}$, we thus have the usual Dirac and Pauli form factors $F_1$ and $F_2$,
\beq
\label{EM_nucleon}
\langle N(p')|j^\mu_\text{em}|N(p)\rangle 
=\bar u(p')\Big[F_1^N(t)\gamma^\mu - F_2^N(t)\frac{i\sigma^{\mu\nu}q_\nu}{2\mN}\Big]u(p),
\eeq
where $j^\mu_\text{em}=\sum_{q=u,d,s} \bar q\mathcal{Q}_q\gamma^\mu q$, $\mathcal{Q}=\text{diag}(2,-1,-1)/3$, and $q=p-p'$. For small momentum transfer $t=(p'-p)^2$, it is sufficient to consider the expansion around $t=0$ 
\begin{align}
\label{FF_expansion}
F_1^N(t)&=Q^N +\frac{\langle r_1^2\rangle^N}{6} t+\Order(t^2),\notag\\
F_2^N(t)&=\kappa^N+\Order(t),
\end{align}
with charge $Q^N$, anomalous magnetic moment $\kappa^N$, and 
\beq
\langle r_1^2\rangle^N=\langle r_E^2\rangle^N-\frac{3\kappa_N}{2\mN^2},
\eeq
in terms of the charge radius $\langle r_E^2\rangle^N$. We will use the numerical values given in Table~\ref{tab:matrix_elements}. In particular, we will use the proton charge radius from muonic atoms $\sqrt{\langle r_E^2\rangle^p}=0.84087(39)\fm$~\cite{Pohl:2010zza,Antognini:1900ns}, in line with most recent electron spectroscopy measurements~\cite{Beyer:2017gug,Fleurbaey:2018fih,Bezginov:2019mdi}, the PRad electron scattering data~\cite{Xiong:2019umf}, and the expectation from dispersion relations~\cite{Lorenz:2012tm,Hoferichter:2016duk}. For the neutron, we use the charge radius from Ref.~\cite{Tanabashi:2018oca}, but note that a recent extraction from the deuteron points to a slightly smaller value~\cite{Filin:2019eoe}. 

\subsection{Dimension-$\boldsymbol{6}$ matrix elements}

At dimension $6$ we first need the vector matrix elements for each quark flavor separately
\begin{align}
\label{vector_FF_def}
&\langle N(p')|\bar q \gamma^\mu q|N(p)\rangle \notag\\
&=\bar u(p')\Big[F_1^{q,N}(t)\gamma^\mu - F_2^{q,N}(t)\frac{i\sigma^{\mu\nu}q_\nu}{2\mN}\Big]u(p).
\end{align}
To perform the flavor decomposition, we will assume isospin symmetry (see Ref.~\cite{Kubis:2006cy} for corrections), which leads to
\begin{align}
\label{vector_FF}
 F_i^{u,p}(t)&=F_i^{d,n}(t)=2F_i^p(t)+F_i^n(t)+F_i^{s,N}(t),\notag\\
 F_i^{u,d}(t)&=F_i^{u,n}(t)=F_i^p(t)+2F_i^n(t)+F_i^{s,N}(t),\notag\\
 F_i^{s,p}(t)&=F_i^{s,n}(t)\equiv F_i^{s,N}(t).
\end{align}
Information on the strangeness form factors has traditionally been extracted from PVES, but the uncertainties are sizable~\cite{Gonzalez-Jimenez:2014bia}. More recently, it has been shown in lattice QCD that the strangeness contribution is very small, in Table~\ref{tab:matrix_elements} we quote the naive average of 
Refs.~\cite{Djukanovic:2019jtp,Alexandrou:2019olr}. 

\begin{table}[t]
	\centering
	\renewcommand{\arraystretch}{1.3}
	\begin{tabular}{lrr}
		\toprule
		$\kappa^p$ & $1.79284734462(82)$ & \cite{Tanabashi:2018oca,Schneider:2017lff}\\
		$\kappa^n$ & $-1.91304273(45)$ & \cite{Tanabashi:2018oca,Mohr:2015ccw}\\
		$\langle r_E^2\rangle^p$ $[\fm^2]$& $0.7071(7)$ & \cite{Antognini:1900ns}\\
		$\langle r_E^2\rangle^n$ $[\fm^2]$& $-0.1161(22)$ & \cite{Tanabashi:2018oca,Koester:1995nx,Kopecky:1997rw}\\
		$\kappa_s^N$ & $-0.017(4)$ & \cite{Djukanovic:2019jtp,Alexandrou:2019olr} \\
		$\langle r_{E,s}^2\rangle^N$ $[\fm^2]$& $-0.0048(6)$ &\cite{Djukanovic:2019jtp,Alexandrou:2019olr}\\\colrule
		$g_A$ & $1.27641(56)$ & \cite{Markisch:2018ndu}\\
		$g_A^{u,p}$ & $0.842(12)$ & \cite{Tanabashi:2018oca,Airapetian:2006vy}\\
		$g_A^{d,p}$ & $-0.427(13)$ & \cite{Tanabashi:2018oca,Airapetian:2006vy}\\
		$g_A^{s,N}$ & $-0.085(18)$ & \cite{Tanabashi:2018oca,Airapetian:2006vy}\\
		$\langle r_A^2\rangle$ $[\fm^2]$ & $0.46(16)$ &\cite{Hill:2017wgb}\\\colrule
		$F_{1,T}^{u,p}(0)$ & $0.784(28)$ & \cite{Gupta:2018lvp}\\
		$F_{1,T}^{d,p}(0)$ & $-0.204(11)$ & \cite{Gupta:2018lvp}\\
		$F_{1,T}^{s,N}(0)$ & $-0.0027(16)$ & \cite{Gupta:2018lvp}\\
		$F_{2,T}^{u,p}(0)$ & $-1.5(1.0)$ & \cite{Hoferichter:2018zwu}\\
		$F_{2,T}^{d,p}(0)$ & $0.5(3)$ & \cite{Hoferichter:2018zwu}\\
		$F_{2,T}^{s,N}(0)$ & $0.009(5)$ & \cite{Hoferichter:2018zwu}\\
		$F_{3,T}^{u,p}(0)$ & $0.1(2)$ & \cite{Hoferichter:2018zwu}\\
		$F_{3,T}^{d,p}(0)$ & $-0.6(3)$ & \cite{Hoferichter:2018zwu}\\
		$F_{3,T}^{s,N}(0)$ & $-0.004(3)$ & \cite{Hoferichter:2018zwu}\\\colrule
		$f_u^p$ $[10^{-3}]$ & $20.8(1.5)$ & \cite{Hoferichter:2015dsa}\\
		$f_d^p$ $[10^{-3}]$ & $41.1(2.8)$ & \cite{Hoferichter:2015dsa}\\
		$f_u^n$ $[10^{-3}]$  & $18.9(1.4)$ & \cite{Hoferichter:2015dsa}\\
		$f_d^n$ $[10^{-3}]$ & $45.1(2.7)$ & \cite{Hoferichter:2015dsa}\\
		$f_s^N$ $[10^{-3}]$ & $43(20)$ & \cite{Durr:2015dna,Yang:2015uis,Abdel-Rehim:2016won,Bali:2016lvx}\\
		$f_Q^N$ $[10^{-3}]$ & $68(1)$ & \cite{Hoferichter:2017olk,Hill:2014yxa}\\
		$\dot\sigma$ $[\GeV^{-1}]$ & $0.27(1)$ &\cite{Hoferichter:2016nvd,Hoferichter:2012wf,Ditsche:2012fv}\\
		$\dot\sigma_s$ $[\GeV^{-1}]$ & $0.3(2)$ &\cite{Hoferichter:2016nvd,Hoferichter:2012wf,Ditsche:2012fv}\\
		$f_u^\pi$ & $0.315(14)$ & \cite{Hoferichter:2015ipa,Fodor:2016bgu}\\
		$f_d^\pi$ & $0.685(14)$ & \cite{Hoferichter:2015ipa,Fodor:2016bgu}\\
		\botrule
	\end{tabular}
	\renewcommand{\arraystretch}{1.0}
	\caption{Values of the hadronic matrix elements.}
	\label{tab:matrix_elements} 
\end{table}

The second dimension-$6$ operator requires input on the axial-vector form factors, as they appear in the decomposition
\begin{align}
\label{axial_vector_FF_def}
&\langle N(p')|\bar q \gamma^\mu\gamma_5 q|N(p)\rangle\notag\\
&= \bar u(p')\Big[\gamma^\mu\gamma_5 G_A^{q,N}(t)-\gamma_5\frac{q^\mu}{2\mN} G_P^{q,N}(t)\notag\\
&\qquad-\frac{i\sigma^{\mu\nu}}{2\mN}q_\nu\gamma_5 G_T^{q,N}(t)\Big]u(p),
\end{align}
where, for completeness, we included the second-class current $G_T^{q,N}(t)$~\cite{Weinberg:1958ut}, but will not further consider its contribution in the following. The normalization is determined by the axial-vector charges
\beq
G_A^{q,N}(0)\equiv g_A^{q,N} \equiv \Delta q^N. 
\eeq
Assuming again isospin symmetry
\beq
\label{gAiso}
g_A^{u,p}=g_A^{d,n},\qquad g_A^{d,p}=g_A^{u,n},\qquad g_A^{s,p}=g_A^{s,n}\equiv g_A^{s,N},
\eeq
these couplings are constrained by
\beq
\label{gA}
g_A^{u,p}-g_A^{d,p}=g_A,\qquad g_A^{u,N}+g_A^{d,N}-2g_A^{s,N}=3F-D,
\eeq
in terms of the axial-vector coupling of the nucleon $g_A=1.27641(56)$~\cite{Markisch:2018ndu} and the $SU(3)$ couplings $D$ and $F$ that can be extracted from semileptonic hyperon decays. In combination with the singlet combination from Ref.~\cite{Airapetian:2006vy}, this leads to the couplings listed in Table~\ref{tab:matrix_elements}. These values are in reasonable agreement with lattice QCD~\cite{Aoki:2019cca} 
\begin{align}
&N_f=2+1+1\,\text{\cite{Lin:2018obj}}: g_A^{u,p}=0.777(39),\notag\\
&\qquad g_A^{d,p}=-0.438(35),\quad g_A^{s,N}=-0.053(8),\notag\\
&N_f=2+1\,\text{\cite{Liang:2018pis}}: g_A^{u,p}=0.847(37),\notag\\
&\qquad g_A^{d,p}=-0.407(24),\quad g_A^{s,N}=-0.035(9),
\end{align}  
but in view of the present uncertainties we adopt the phenomenological determination. However, while part of the difference to phenomenology could be due to the scale dependence of the singlet combination,
both lattice calculations point to a smaller strangeness coupling than extracted from the spin structure functions.   

The triplet and octet components of the induced pseudoscalar form factor $G_P(t)$ are constrained by Ward identities, whose manifestation at leading order in the chiral expansion becomes
\begin{align}
\label{GA_LO}
G_A^3(t)&=g_A,\qquad G_A^8(t)=\frac{3F-D}{\sqrt{3}}\equiv g_A^8,\notag\\
G_P^3(t)&=-\frac{4\mN^2 g_A}{t-\mpi^2},\qquad G_P^8(t)=-\frac{4\mN^2 g_A^8}{t-\meta^2}.
\end{align}
Finally, for the triplet component there is also experimental information on the momentum dependence. Defining the axial radius by
\beq
\label{axial_radius}
G_A^3(t)=g_A\bigg(1+\frac{\langle r^2_A\rangle}{6} t+\Order(t^2)\bigg),
\eeq
a simple dipole ansatz 
\beq
G_A^3(t)=\frac{g_A}{(1-t/M_A^2)^2},
\eeq
with mass scale $M_A$ around $1\GeV$~\cite{Bodek:2007ym}, implies 
$\langle r_A^2\rangle = 12/M_A^2\sim 0.47\fm^2$. The central value agrees with Ref.~\cite{Hill:2017wgb}, a global analysis of muon capture and neutrino scattering, but the uncertainties are substantial, see Table~\ref{tab:matrix_elements}. 
To ensure that the Ward identity is satisfied up to higher orders, the pseudoscalar form factor needs to be modified according to~\cite{Bernard:2001rs}
\beq
G_P^3(t)=-\frac{4\mN g_{\pi NN}F_\pi}{t-\mpi^2} -\frac{2}{3}g_A \mN^2 \langle r^2_A\rangle+\Order\big(t,\mpi^2\big)
\eeq
when including the radius corrections~\eqref{axial_radius}.
The full $\pi N$ coupling constant $g_{\pi NN}$ has been introduced as a convenient way to capture all chiral corrections at $\Order(1)$. In the numerical analysis we will use $F_\pi=92.28(9)\MeV$~\cite{Tanabashi:2018oca} and $g_{\pi NN}^2/(4\pi)=13.7(2)$~\cite{deSwart:1997ep,Baru:2010xn,Baru:2011bw,Perez:2016aol,Reinert:2020mcu}. With this input, the Goldberger--Treiman discrepancy becomes
\beq
\Delta_\text{GT}=\frac{g_{\pi NN} F_\pi}{\mN g_A}-1=1.0(7)\%,
\eeq
demonstrating that the chiral corrections are rather small.

The final matrix elements in the dimension-$6$ Lagrangian concern the tensor operator $\bar q\sigma^{\mu\nu} q$, for which 
we use the decomposition
\begin{align}
\label{tensor_decomposition}
 &\langle N(p')|\bar q \sigma^{\mu\nu} q|N(p)\rangle\notag\\
 &=\bar u(p')\Big[\sigma^{\mu\nu} F_{1,T}^{q,N}(t) - \frac{i}{\mN}\big(\gamma^\mu q^\nu-\gamma^\nu q^\mu\big)F^{q,N}_{2,T}(t)\notag\\
 &\qquad-\frac{i}{\mN^2}\big(P^\mu q^\nu-P^\nu q^\mu\big)F^{q,N}_{3,T}(t)\Big]u(p).
\end{align}
The tensor charges $g_T^{q,p}=F_{1,T}^{q,p}(0)$, as given in Table~\ref{tab:matrix_elements}, are taken from lattice QCD~\cite{Gupta:2018lvp}. The other form-factor normalizations come from Ref.~\cite{Hoferichter:2018zwu} (with strangeness input updated to Ref.~\cite{Alexandrou:2019olr}).

\subsection{Dimension-$\boldsymbol{7}$ matrix elements}

At dimension $7$ we first need the scalar matrix elements of the nucleon
\beq
\label{scalar_current}
\langle N(p')|m_q \bar q q|N(p)\rangle = \mN f_q^N(t) \bar u(p') u(p).
\eeq
To separate the momentum dependence we define
\begin{align}
f_u^N(t)&= f_u^N+\frac{1-\xi_{ud}}{2\mN}\dot\sigma t+\Order(t^2),\notag\\
f_d^N(t)&= f_d^N+\frac{1+\xi_{ud}}{2\mN}\dot\sigma t+\Order(t^2),\notag\\
f_s^N(t)&= f_s^N+\frac{\dot\sigma_s}{\mN} t+\Order(t^2).
\end{align}
The scalar couplings $f_{u,d}^N$ are largely determined by the pion--nucleon $\sigma$-term $\sigma_{\pi N}$~\cite{Crivellin:2013ipa}, up to isospin-breaking corrections that can be extracted from the proton--neutron mass difference~\cite{Gasser:1974wd,Borsanyi:2014jba,Gasser:2015dwa,Brantley:2016our,Gasser:2020mzy}. The numbers given in Table~\ref{tab:matrix_elements} follow from $\sigma_{\pi N}$ as extracted from data on pionic atoms~\cite{Gotta:2008zza,Strauch:2010vu,Hennebach:2014lsa,Baru:2010xn,Baru:2011bw} when used as input for a dispersive analysis of pion--nucleon scattering~\cite{Hoferichter:2015dsa,Hoferichter:2015hva}. This result has been confirmed using independent input from scattering data~\cite{RuizdeElvira:2017stg}, but there is a persistent tension with lattice QCD that still has not been resolved~\cite{Durr:2015dna,Yang:2015uis,Abdel-Rehim:2016won,Bali:2016lvx,Hoferichter:2016ocj,Borsanyi:2020}. Accordingly, we have increased the error in $f_s^N$ given that in this case all phenomenological extractions are subject to large $SU(3)$ uncertainties.  
The heavy-quark couplings $f_Q$ effectively describe the matrix element of the trace anomaly at $\Order(\alpha_s)$
\begin{align}
f_Q^N(t)&=\frac{2}{27}\bigg(\frac{\theta_0^N(t)}{\mN}-\sum_{q=u,d,s}f_q^N(t)\bigg),\notag\\
\theta_0^N(t)&=\langle N(p')|\theta^\mu_\mu|N(p)\rangle,
\end{align}
with normalization $\theta_0^N(0)=\mN$, while perturbative corrections especially for the charm quark lead to additional uncertainties~\cite{Hill:2014yxa}.
For the momentum dependence, $\dot\sigma$ and $\dot \sigma_s$ are taken from 
Refs.~\cite{Hoferichter:2016nvd,Hoferichter:2012wf,Ditsche:2012fv}, 
and~\cite{Aoki:2019cca,Fodor:2016bgu}
\beq
\xi_{ud}=\frac{m_d-m_u}{m_d+m_u}=0.35(2),
\eeq 
which also determines the scalar couplings of the pion
\beq
\langle\pi|m_q \bar q q|\pi\rangle = f_q^\pi \mpi^2,
\eeq
according to
\begin{align}
f_u^\pi&=\frac{m_u}{m_u+m_d}=\frac{1}{2}\big(1-\xi_{ud}\big)=0.315(14),\notag\\
f_d^\pi&=\frac{m_d}{m_u+m_d}=\frac{1}{2}\big(1+\xi_{ud}\big)=0.685(14).
\end{align}
These matrix elements arise in two-body corrections to scalar currents and are also included in Table~\ref{tab:matrix_elements}.

Finally, we parameterize the pseudoscalar matrix element as
\beq
\langle N(p')|m_q\bar q i\gamma_5 q|N(p)\rangle = \mN G_5^{q,N}(t) \bar u(p')i\gamma_5 u(p).
\eeq
For the nonsinglet component the new form factor is related to the axial-vector ones by the Ward identity
\beq
G_5^{q,N}(t)=G_A^{q,N}(t)+\frac{t}{4\mN^2}G_P^{q,N}(t),
\eeq
but in the singlet case this relation is broken by the anomaly contribution from $G_{\mu\nu}^a\tilde G^{\mu\nu}_a$, similarly to the gluonic contribution to the trace anomaly. For a consistent treatment of singlet effects one would thus have to extend the operator basis in Eq.~\eqref{operators_basis} accordingly. In the past, the singlet pseudoscalar matrix element has often been estimated by assuming~\cite{Cheng:1988im,Cheng:2012qr} 
\beq
\langle N|\sum_{q=u,d,s} \bar q i\gamma_5 q|N\rangle=0,
\eeq
but the analogous relation for the axial-vector singlet combination $\sum_{q=u,d,s}\Delta q$ does not display the expected $1/N_c$ suppression. 
The matrix element of the gluon anomaly could be extracted with similar techniques as used for lattice calculations of the QCD $\theta$ term~\cite{Dragos:2019oxn}.

\subsection{Nuclear responses}
\label{sec:nuclear_responses}

\begin{table}[t]
	\centering
	\renewcommand{\arraystretch}{1.3}
	\begin{tabular}{lrrr}
		\toprule
		responses & operator & coherence & interpretation\\\colrule
		$\F^M_\pm$ & $\unity$ & coherent & charge\\
		$\F^{\Phi''}_\pm$ & $\spin_N\cdot (\qq\times \PP)$ & semi-coherent & spin-orbit\\
		$S_{ij}$ & $\spin_N$ & not coherent & axial\\ 
		\botrule
	\end{tabular}
	\renewcommand{\arraystretch}{1.0}
	\caption{Nomenclature for the nuclear structure factors. The second column gives the leading operators on the single-nucleon level, the third one indicates the extent to which the response scales coherently with nucleon number, and the fourth column gives its physical interpretation. The axial responses include longitudinal, transverse electric, and transverse magnetic multipoles. $\spin_N=\sig_N/2$ denotes the nucleon spin operator and the momenta are defined as in Sec.~\ref{sec:SM}.}
	\label{tab:nuclear_structure_factors} 
\end{table}

As a final step, the nucleon-level matrix elements need to be convolved with the nuclear states. Formally, the decomposition into distinct nuclear responses requires a multipole decomposition, see Refs.~\cite{Serot:1978vj,Donnelly:1978tz,Donnelly:1979ezn,Serot:1979yk,Donnelly:1984rg,Walecka:1995mi}, which in full generality becomes very complex. 
Here, we concentrate on the most relevant contributions, with the main features summarized in Table~\ref{tab:nuclear_structure_factors}, and review some of the details needed later in App.~\ref{app:multipole}.

By far the most important response is related to the charge operator, it is denoted by the structure factors $\F^M_\pm(\qq^2)$ normalized by
\beq
\F^M_+(0)=N+Z=A,\qquad \F^M_-(0)=Z-N.
\eeq
This is the only response that is fully coherent. In addition to $\F^M_\pm$, we also need the so-called $\F^{\Phi''}_\pm$ structure factor, which can be interpreted in terms of spin-orbit corrections. This response vanishes for $\qq^2=0$, but it interferes with 
$\F^M_\pm$ and receives some coherent enhancement, especially for heavy nuclei.
This is because in the relevant nuclei nucleons tend to occupy orbitals with spin parallel to the angular orbital momentum (lowered in energy by the nuclear spin-orbit interaction) and high-energy orbitals with antiparallel spin, which would cancel $\F^{\Phi''}_\pm$, remain mostly empty. The interference with $\F^M$ and partial coherence make the $\Phi''$ response the most relevant correction. In principle, both $\F^M_\pm$, $\F^{\Phi''}_\pm$ may contribute beyond the leading $L=0$ multipole, but such effects are not coherent, vanish at $\qq^2=0$, and without interference with the leading multipole effectively become negligible. Due to this we will continue to identify both responses with their $L=0$ multipole.

Finally, there are several responses that emerge from the axial-vector operator. Their contribution again is not coherent, but remains finite at $\qq^2=0$. In these cases, several multipoles and responses become relevant, but we will continue to use a notation in which these effects are subsumed into structure factors $S_{ij}$ (with indices $i,j=0,1$ corresponding to isoscalar/isovector combinations). 
We keep the induced pseudoscalar form factors $G_P^{q,N}$, whose contribution
is enhanced by the presence of the pion pole, but do not consider any other subleading noncoherent responses. 
Further aspects of the multipole decomposition are discussed in Sec.~\ref{sec:SM} whenever necessary to introduce the nuclear responses for a given process.

\section{Nuclear responses in the Standard Model}
\label{sec:SM}

In this section we will collect the nuclear responses as they appear in electron--nucleus and neutrino--nucleus scattering. In particular, we demonstrate how the traditional charge and weak form factors emerge in the formalism established in Sec.~\ref{sec:formalism}. In either case, the kinematics are defined by
\beq
\ell(k)+\N(p) \to \ell(k')+\N(p'),\qquad \ell\in\{e^-,\nu\},
\eeq
with 
\beq
q=k'-k=p-p',
\label{defq}
\eeq
and invariants
\beq
\label{Mandelstam}
s=(p+k)^2,\qquad t=(p-p')^2,\qquad u=(p-k')^2,
\eeq
fulfilling $s+t+u=2\mA^2$. Here, $\mA$ refers to the mass of the nucleus and lepton masses are neglected throughout. We also define $P=p+p'$ and write $t=q^2=-Q^2$.

\subsection{Parity-conserving electron scattering}

For electron scattering, the invariants~\eqref{Mandelstam} are conventionally replaced in favor of
\beq
\eta=-\frac{t}{4\mA^2},\qquad z=\cos\theta=1-\frac{2\mA^2 t}{(s-\mA^2)(u-\mA^2)},
\eeq
where $\theta$ is the scattering angle in the laboratory frame. In this frame the relation of the spin-averaged scattering amplitude $|\bar \M|^2$  to the cross section becomes
\begin{align}
\frac{\diff\sigma}{\diff\Omega}&=\frac{|\bar \M|^2}{64\pi^2\mA(\mA+E(1-z))}\frac{E'}{E}\notag\\
&=\bigg(\frac{\diff\sigma}{\diff\Omega}\bigg)_\text{Mott}\times \frac{E'}{E}\times\frac{t^2|\bar \M|^2}{4e^4(\mA^4-s u)},
\end{align}
where the last relation defines the Mott cross section
\begin{align}
\bigg(\frac{\diff\sigma}{\diff\Omega}\bigg)_\text{Mott}&=\frac{e^4(\mA^4-s u)}{16\pi^2t^2\mA(\mA+E(1-z))}\notag\\
&=\frac{\alpha^2}{4E^2}\frac{\cos^2\frac{\theta}{2}}{\sin^4\frac{\theta}{2}}.
\end{align}
The incoming and outgoing electron energies are given by
\beq
E=\frac{s-\mA^2}{2\mA},\qquad E'=\frac{\mA^2-u}{2\mA}.
\eeq

For the parity-conserving case, the amplitude becomes
\begin{align}
 |\bar \M|^2&=\frac{1}{2(2J+1)}\sum_\text{spins}|\M|^2,\notag\\
 \M&=-\frac{e^2}{t}\bar u (k')\gamma_\mu u(k)\,\langle \N(p')|j^\mu_\text{em}|\N(p)\rangle,
\end{align}
and at the single-nucleon level the hadronization follows from Eq.~\eqref{EM_nucleon}.
The leptonic trace
\beq
L^{\mu\nu}=\Tr\big(\slashed{k'}\gamma^\mu\slashed{k}\gamma^\nu)
=4\big(k^\mu k'^\nu+k'^\mu k^\nu-g^{\mu\nu}k\cdot k'\big),
\eeq
fulfilling $k_\mu L^{\mu\nu}=k'_\mu L^{\mu\nu}=0$, needs to be contracted with the nuclear amplitude, which we express in terms of multipoles according to Sec.~\ref{sec:nuclear_responses}, see Ref.~\cite{Donnelly:1984rg} and App.~\ref{app:multipole}. The leading terms can be read off from the nonrelativistic expansion of the single-nucleon current
\begin{align}
\label{EM_NR}
\langle N(p')|j^0_\text{em}|N(p)\rangle&= F_1^N(t)+\frac{F_1^N(t)+2F_2^N(t)}{8\mN^2}t\\
&-i \spin_N\cdot (\qq\times \PP) \frac{F_1^N(t)+2F_2^N(t)}{4\mN^2},\notag
\end{align}
where we dropped the remaining two-component spinors and the space-like components do not contribute to the $M$ and $\Phi''$ responses.  
After the multipole decomposition, the first line of Eq.~\eqref{EM_NR} will contribute to $\F^M$, the second to $\F^{\Phi''}$, and the combination to an interference term between the two responses. Concentrating on the $L=0$ multipole, the result takes a very compact form and is typically expressed as
\beq
\label{parity_conserving_scattering}
\frac{\diff\sigma}{\diff\Omega}=\bigg(\frac{\diff\sigma}{\diff\Omega}\bigg)_\text{Mott}\times \frac{E'}{E}\times
Z^2 \times \big[F_\text{ch}(\qq^2)\big]^2,
\eeq
with the charge form factor defined by
\begin{align}
\label{Fcharge}
 F_\text{ch}(\qq^2)&=\frac{1}{Z}\bigg[\bigg(1+\frac{\langle r_E^2\rangle^p}{6}t 
 + \frac{1}{8\mN^2} t\bigg) \F^M_p(\qq^2)\notag\\
 &+\frac{\langle r_E^2\rangle^n}{6} t \F^M_n(\qq^2)\notag\\
&- \frac{1+2\kappa_p}{4\mN^2} t \F^{\Phi''}_p(\qq^2)
- \frac{2\kappa_n}{4\mN^2} t \F^{\Phi''}_n(\qq^2)\bigg].
\end{align}
The proton/neutron combinations are related to the isospin components by
\begin{align}
\label{Fpn}
\F^M_\pm(\qq^2)&=\F^M_p(\qq^2)\pm \F^M_n(\qq^2),\notag\\
\F^{\Phi''}_\pm(\qq^2)&=\F^{\Phi''}_p(\qq^2)\pm \F^{\Phi''}_n(\qq^2),
\end{align}
and we have replaced the full form factors in Eq.~\eqref{EM_NR} by the first terms in the expansion~\eqref{FF_expansion}.
The charge form factor fulfills the normalization $F_\text{ch}(0)=1$, and the corresponding representation~\eqref{parity_conserving_scattering} is exact for spin-$0$ nuclei. For nonvanishing spin, there are further form factors, e.g., the magnetic form factor for spin-$1/2$ in analogy to the nucleon, but for the reasons given in Sec.~\ref{sec:nuclear_responses} these contributions are small in heavy nuclei. In addition, two-body effects only enter at loop level, so that in contrast to the magnetic form factor two-body modifications of the charge form factor are also small.
Finally, we give the corresponding expansion for the charge radius
\begin{align}
 R_\text{ch}^2&=R_p^2 + \langle r_E^2\rangle^p + \frac{N}{Z} \langle r_E^2\rangle^n + \frac{3}{4\mN^2}+\langle r^2\rangle_\text{so},\\
 \langle r^2\rangle_\text{so}&=-\frac{3}{2\mN^2Z}\Big(\big(1+2\kappa_p\big) \F^{\Phi''}_p(0)
+2\kappa_n\F^{\Phi''}_n(0)\Big),\notag
\end{align}
where $R_p^2$ is the so-called point-proton radius defined as 
\beq
R_p^2=-\frac{6}{Z}\frac{\diff\F^M_p(\qq^2)}{\diff\qq^2}\bigg|_{\qq^2=0},
\eeq
and $\langle r^2\rangle_\text{so}$ represents the spin-orbit contribution encoded in $\Phi''$~\cite{Bertozzi:1972jff}.
In the case of Eq.~\eqref{Fcharge} the matching of matrix elements and Wilson coefficients is trivial, since so far only the long-range contribution in the SM has been taken into account. A potential modification would be provided by the electron analog of $\Lagr^{(5)}$ given in Eq.~\eqref{operators_basis}. In the next step, we extend the discussion to short-range contributions from $Z$ exchange, which are responsible for PVES in the SM.

\subsection{Parity-violating electron scattering}
\label{sec:PVES}

The central observable in PVES is the asymmetry
\beq
A_\text{PVES}=\frac{\big(\frac{\diff\sigma}{\diff\Omega}\big)_R-\big(\frac{\diff\sigma}{\diff\Omega}\big)_L}{\big(\frac{\diff\sigma}{\diff\Omega}\big)_R+\big(\frac{\diff\sigma}{\diff\Omega}\big)_L},
\eeq
where the cross sections involve left- or right-handed initial-state electrons, respectively. The corresponding lepton traces are   
\begin{align}
\label{PVES_lepton}
L^{\mu\nu}_{L/R}&=\Tr\big(\slashed{k'}\gamma^\mu P_{L/R}\slashed{k}\gamma^\nu(g_V^e-g_A^e\gamma_5)\big)\\
&=2(g_V^e\pm g_A^e)\notag\\
&\qquad\times\big(k^\mu k'^\nu+k'^\mu k^\nu-g^{\mu\nu}k\cdot k'\pm i\eps^{\mu\nu\alpha\beta}k_\alpha k'_\beta\big),\notag
\end{align}
where
\beq
g_V^e=-\frac{1}{2}+2\sin^2\theta_W,\qquad g_A^e=-\frac{1}{2},
\eeq
are the vector and axial-vector weak charges of the electron in the normalization of Ref.~\cite{Tanabashi:2018oca}.
The terms in Eq.~\eqref{PVES_lepton} involving an $\eps$ tensor will lead to SD corrections, which we will study below in the context of CE$\nu$NS, while the remainder follows in close analogy to the parity-conserving case, the only difference being that the electromagnetic form factor needs to be replaced by its weak analog. With quark-level Wilson coefficients as in the SM and matrix elements from Eq.~\eqref{vector_FF}, the result takes the simple form
\beq
A_\text{PVES}=\frac{G_F t}{4\pi\alpha\sqrt{2}}\frac{Q_\text{w} F_\text{w}(\qq^2)}{Z F_\text{ch}(\qq^2)},
\eeq
where the weak charge 
\begin{align}
\label{Qweak}
Q_\text{w}&=ZQ_\text{w}^p+NQ_\text{w}^n,\notag\\
Q_\text{w}^p&=1-4\sin^2\theta_W,\qquad Q_\text{w}^n=-1,
\end{align}
has been separated from the weak form factor $F_\text{w}(\qq^2)$. However, we note that, in contrast to the electromagnetic charge and $F_\text{ch}(\qq^2)$, $Q_\text{w}$ does not actually factorize. The explicit definition reads
\begin{align}
\label{Fw_definition}
 F_\text{w}(\qq^2)&=\frac{1}{Q_\text{w}}\bigg[\bigg(Q_\text{w}^p\Big(1+\frac{\langle r_E^2\rangle^p}{6} t +\frac{1}{8\mN^2} t\Big)\\
 &\qquad+Q_\text{w}^n \frac{\langle r_E^2\rangle^n+\langle r_{E,s}^2\rangle^N}{6} t\bigg) \F^M_p(\qq^2)\notag\\
&+\bigg(Q_\text{w}^n\Big(1+\frac{\langle r_E^2\rangle^p+\langle r_{E,s}^2\rangle^N}{6} t +\frac{1}{8\mN^2} t\Big)\notag\\
 &\qquad+Q_\text{w}^p \frac{\langle r_E^2\rangle^n}{6} t\bigg)\F^M_n(\qq^2)\notag\\
&- \frac{Q_\text{w}^p(1+2\kappa^p)+2Q_\text{w}^n(\kappa^n+\kappa_s^N)}{4\mN^2} t \F^{\Phi''}_p(\qq^2)\notag\\
&- \frac{Q_\text{w}^n(1+2\kappa^p+2\kappa_s^N)+2Q_\text{w}^p\kappa^n}{4\mN^2} t \F^{\Phi''}_n(\qq^2)\bigg],\notag
\end{align}
where we have used that in the SM the Wilson coefficients for $d$- and $s$-quarks are identical to write the strangeness contribution in terms of $Q_\text{w}^n$. The corresponding weak radius reads
\begin{align}
R_\text{w}^2&=\frac{Z Q_\text{w}^p}{Q_\text{w}}\bigg(R_p^2+ \langle r_E^2\rangle^p
+ \frac{Q_\text{w}^n}{Q_\text{w}^p}\big(\langle r_E^2\rangle^n+\langle r_{E,s}^2\rangle^N\big)\bigg)\notag\\
&+\frac{N Q_\text{w}^n}{Q_\text{w}}\bigg(R_n^2+ \langle r_E^2\rangle^p+\langle r_{E,s}^2\rangle^N +  \frac{Q_\text{w}^p}{Q_\text{w}^n}\langle r_E^2\rangle^n\bigg)\notag\\
&+\frac{3}{4\mN^2}+\langle \tilde r^2\rangle_\text{so},
\end{align}
with spin-orbit contribution~\cite{Horowitz:2012we} 
\begin{align}
\langle \tilde r^2\rangle_\text{so}&=-\frac{3Q_\text{w}^p}{2\mN^2Q_\text{w}}\Big(1+2\kappa^p+2\frac{Q_\text{w}^n}{Q_\text{w}^p}(\kappa^n+\kappa_s^N)\Big) \F^{\Phi''}_p(0)\notag\\
&-\frac{3Q_\text{w}^n}{2\mN^2Q_\text{w}}\Big(1+2\kappa^p+2\kappa_s^N+2\frac{Q_\text{w}^p}{Q_\text{w}^n}\kappa^n\Big)\F^{\Phi''}_n(0).
\end{align}
Numerically, we use the values~\cite{Tanabashi:2018oca}
\begin{align}
 Q_\text{w}^p&=-\frac{\sqrt{2}}{G_F}(2C_u^V+C_d^V)=0.0714,\notag\\
 Q_\text{w}^n&=-\frac{\sqrt{2}}{G_F}(C_u^V+2C_d^V)=-0.9900,
\end{align}
and accordingly for $Q_\text{w}$, because the process-dependent radiative corrections~\cite{Tanabashi:2018oca,Erler:2013xha,Blunden:2012ty} as for atomic parity violation or PVES are not yet available.
	
\begin{figure}[t]
	\begin{center}
	\includegraphics[width=0.48\textwidth,clip]{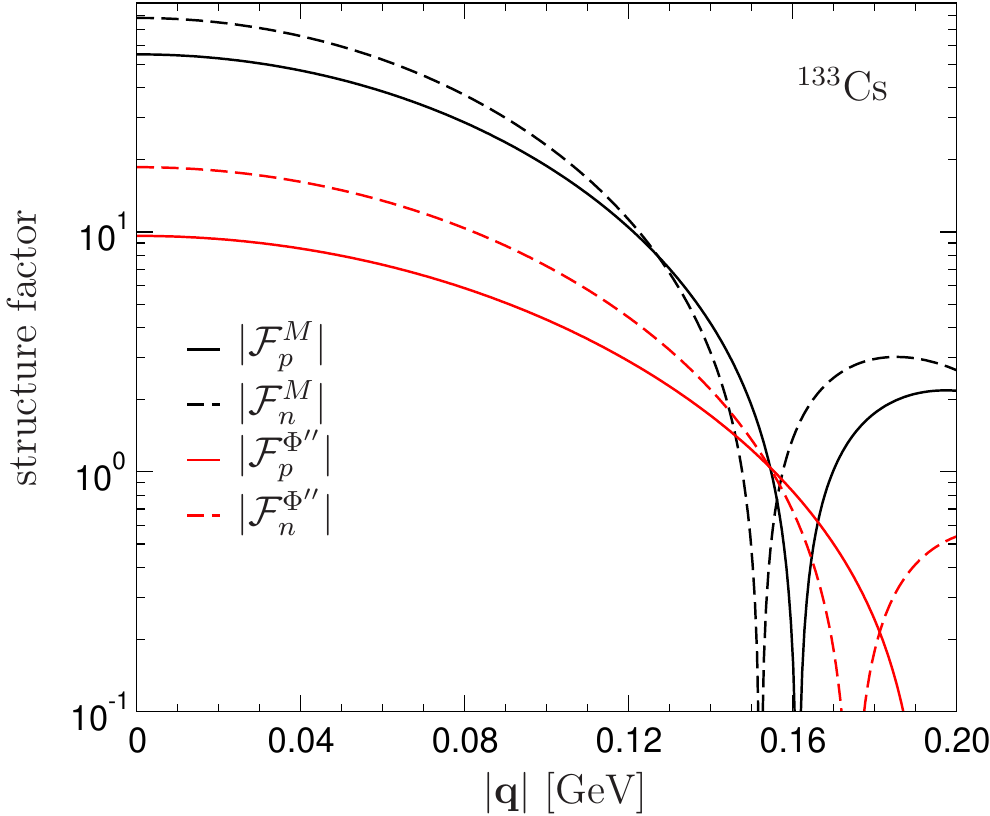}
	\end{center}
\caption{$M$ and $\Phi''$ responses for cesium.}
\label{fig:Cs_M_Phipp}
\end{figure}

We have calculated the nuclear responses $\F^{M}_N(\qq)$, $\F^{\Phi''}_N(\qq)$,
and the corresponding nuclear radii for isotopes relevant for experiment with the nuclear shell model.
The calculations use the same configuration spaces and nuclear interactions as in previous works~\cite{Klos:2013rwa,Hoferichter:2016nvd,Hoferichter:2018acd}.
In particular, the shell-model interactions used are
USDB for $^{19}$F and $^{23}$Na~\cite{Brown:2006gx} (with $0d_{5/2}$, $1s_{1/2}$, and $0d_{3/2}$ single-particle orbitals), SDPF.SM~\cite{Caurier:2007ee} for $^{40}$Ar ($0d_{5/2}$, $1s_{1/2}$, $0d_{3/2}$, $0f_{7/2}$, $1p_{3/2}$, $1p_{1/2}$, and $0f_{5/2}$ space), RG~\cite{Menendez:2009xa} for $^{73}$Ge ($1p_{3/2}$, $0f_{5/2}$, $1p_{3/2}$, and $g_{9/2}$ orbitals), and GCN5082~\cite{Menendez:2008jp} for $^{127}$I, $^{133}$Cs, and $^{129,131}$Xe ($0g_{7/2}$, $1d_{5/2}$, $1s_{1/2}$, $0d_{3/2}$, and $h_{11/2}$ space).
The notation for harmonic oscillator orbitals is $nl_j$, where $n$ is the principal quantum number, and $l,j$ denote the orbital and total angular momentum.
For additional details on the calculations, see Refs.~\cite{Klos:2013rwa,Hoferichter:2016nvd,Hoferichter:2018acd}.
The nuclear-structure calculations have been performed with the
shell-model code ANTOINE~\cite{Caurier:1999,Caurier:2004gf}.

While the phenomenological character of the nuclear interactions used in our work prevents the assessment of reliable nuclear-structure uncertainties, the shell-model results agree very well with experiment.
For instance, our calculations reproduce well the energies of the lowest-lying excited states of these nuclei~\cite{Klos:2013rwa,Vietze:2014vsa,Hoferichter:2018acd} and the electromagnetic transitions between them, including for the ground states involved in CE$\nu$NS~\cite{Hoferichter:2018acd}.
The magnetic moments of the ground states of odd-mass nuclei,
and the quadrupole moments of first excited states are also in good agreement with experiment~\cite{Hoferichter:2018acd}.
From all these isotopes, only $^{133}$Cs is presented here for the first time, compared to our previous works.
This calculation was carried out without truncations in the configuration space,
and the quality of the $^{133}$Cs results is illustrated by the energy spectrum discussed in App.~\ref{app:133cs}.

As an example, Fig.~\ref{fig:Cs_M_Phipp} shows the $M$ and $\Phi''$ responses for $^{133}$Cs.
The coherent and partially coherent characters of $M$ and $\Phi''$, respectively, are clearly observed at $\qq=0$, where about $20\%$--$25\%$ of the nucleons contribute coherently for $\F^{\Phi''}_N(0)$.
The minimum of $\F^{M}_n$ at lower $|\qq|$ compared to $\F^{M}_p$ indicates a larger neutron than proton radius.
Explicit parameterizations of all nuclear structure factors are provided in App.~\ref{app:parameterization}.

\begin{table}[t]
	\centering
	\renewcommand{\arraystretch}{1.3}
	\begin{tabular}{cc|ccccccc}
		\toprule
		& & 	
$^{19}$F & $^{23}$Na & $^{40}$Ar & $^{70}$Ge  \\ 
$R_\text{ch}$ & Th & $2.83$ & $3.01$ & $3.43$ & $4.06$  \\
$R_\text{ch}$ & Exp & $2.8976(25)$ & $2.9936(21)$ & $3.4274(26)$ & $4.0414(12)$  \\
$R_\text{w}$ & Th & $2.90$ & $3.06$ & $3.55$ & $4.14$  \\
$R_n-R_p$ & Th & $0.06$ & $0.04$ & $0.11$ & $0.08$  \\ \colrule
		& & 	
$^{72}$Ge & $^{73}$Ge & $^{74}$Ge & $^{76}$Ge  \\ 
$R_\text{ch}$ & Th & $4.07$ & $4.08$ & $4.08$ & $4.08$  \\
$R_\text{ch}$ & Exp & $4.0576(13)$ & $4.0632(14)$ & $4.0742(12)$ & $4.0811(12)$  \\
$R_\text{w}$ & Th & $4.20$ & $4.23$ & $4.26$ & $4.31$  \\ 
$R_n-R_p$ & Th & $0.13$ & $0.14$ & $0.17$ & $0.21$  \\ \colrule
		& & 	
$^{127}$I & $^{133}$Cs &  &   \\ 
$R_\text{ch}$ & Th & $4.73$ & $4.78$ &  &   \\
$R_\text{ch}$ & Exp & $4.7500(81)$ & $4.8041(46)$ &  &  \\
$R_\text{w}$ & Th & $5.00$ & $5.08$ &  &  \\ 
$R_n-R_p$ & Th & $0.26$ & $0.27$ & &  \\ \colrule
& & 	
$^{128}$Xe & $^{129}$Xe & $^{130}$Xe & $^{131}$Xe  \\ 
$R_\text{ch}$ & Th & $4.75$ & $4.75$ & $4.76$ & $4.77$  \\
$R_\text{ch}$ & Exp & $4.7774(50)$ & $4.7775(50)$ & $4.7818(49)$ & $4.7808(49)$  \\
$R_\text{w}$ & Th & $5.01$ & $5.03$ & $5.04$ & $5.06$  \\ 
$R_n-R_p$ & Th & $0.24$ & $0.26$ & $0.26$ & $0.27$ \\\colrule
& & 	
 $^{132}$Xe & $^{134}$Xe & $^{136}$Xe & \\ 
$R_\text{ch}$ & Th &  $4.77$ & $4.78$ & $4.79$ & \\
$R_\text{ch}$ & Exp &  $4.7859(48)$ & $4.7899(47)$ & $4.7964(47)$ &\\ 
$R_\text{w}$ & Th &  $5.08$ & $5.10$ & $5.13$ & \\ 
$R_n-R_p$ & Th & $0.28$ & $0.30$ & $0.32$ & $ $  \\
		\botrule
	\end{tabular}
	\renewcommand{\arraystretch}{1.0}
	\caption{Shell-model charge and weak radii (in fm). The  experimental data for the charge radii are from Ref.~\cite{Angeli:2013epw}. The table also includes our results for the neutron skin $R_n-R_p$.
	}
	\label{tab:radii}
\end{table} 

We obtain the charge and weak radii given in Table~\ref{tab:radii}.
In addition, Table~\ref{tab:radii} also shows the so-called neutron skin, defined as the difference between neutron and proton point radii, $R_n-R_p$.
Calculated charge radii are in good agreement with experiment, similar to other approaches~\cite{Payne:2019wvy,Co:2020gwl,Cadeddu:2017etk}.
The disagreement between calculations increases for predictions of the weak radii and neutron skin.
The shell model generally predicts larger weak radii and especially larger neutron skins than other many-body approaches~\cite{Payne:2019wvy,Co:2020gwl,Patton:2012jr,Yang:2019pbx,Centelles:2008vu,Cadeddu:2017etk,Brown:2008ib}.

\begin{figure}[t]
	\begin{center}
	\includegraphics[width=0.48\textwidth,clip]{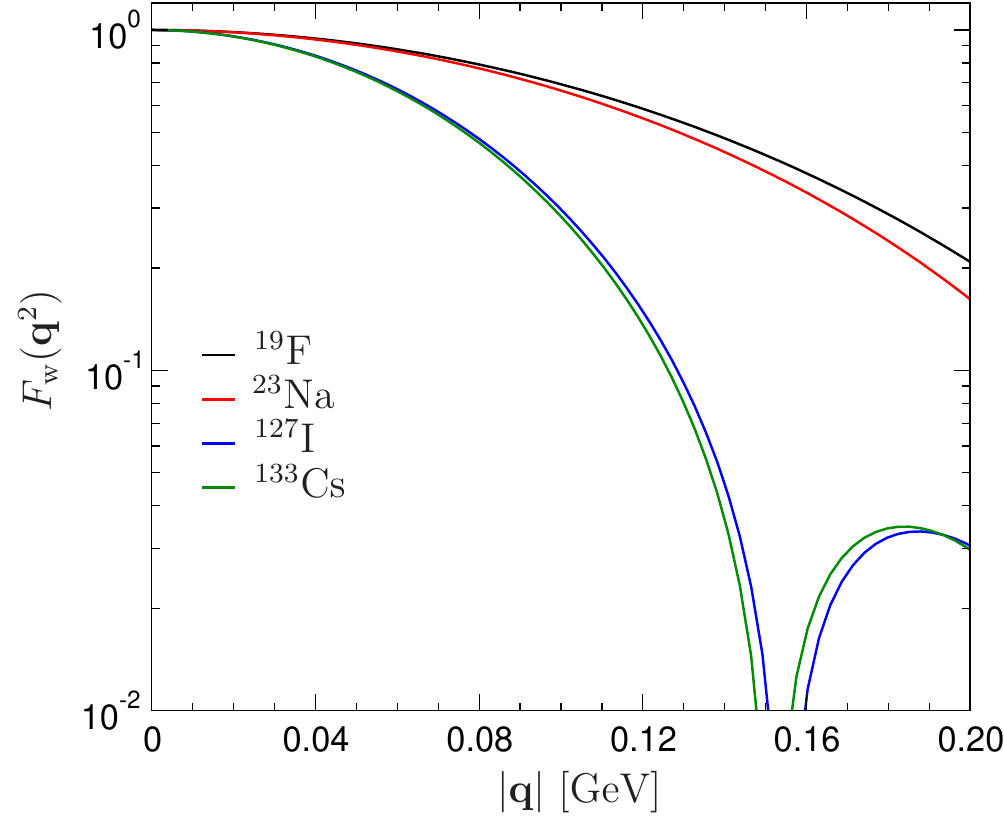}
	\end{center}
\caption{Shell-model results for the weak form factor of $^{19}$F, $^{23}$Na, $^{127}$I, and $^{133}$Cs.}
\label{fig:Fweak_CsI}
\end{figure}

\begin{figure}[t]
	\begin{center}
	\includegraphics[width=0.48\textwidth,clip]{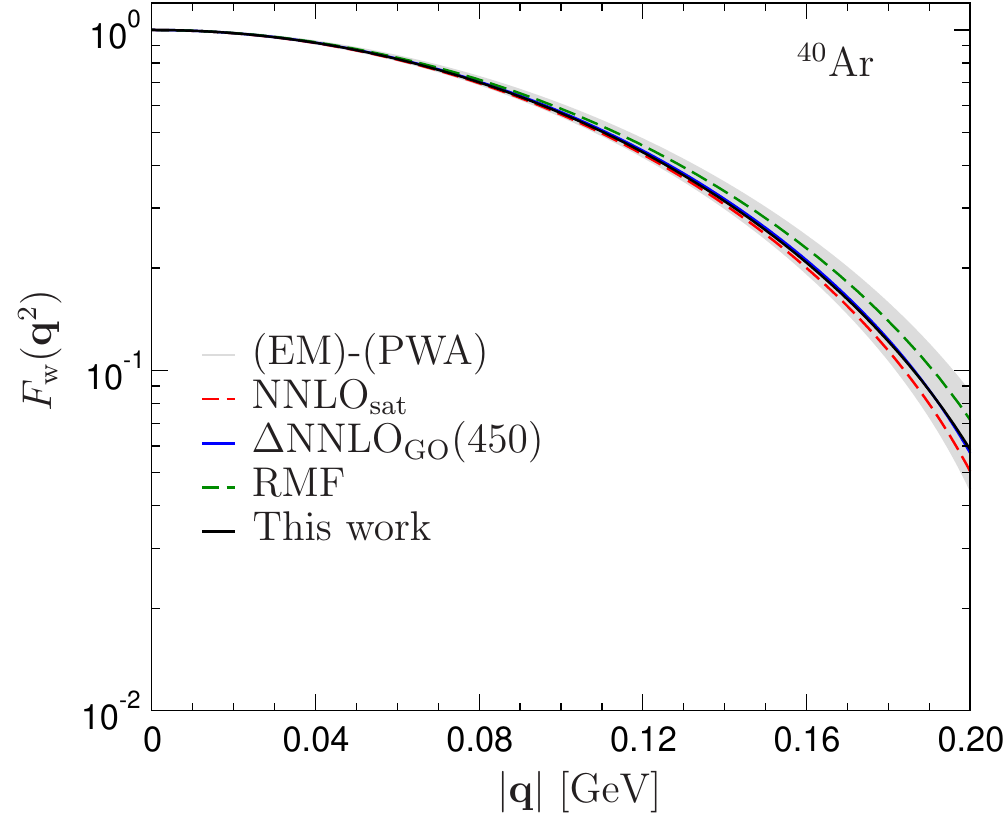}
	\end{center}
\caption{Shell-model results for the weak form factor of $^{40}$Ar, in comparison to RMF~\cite{Yang:2019pbx} and coupled-cluster~\cite{Payne:2019wvy} results. The curves/bands labeled (EM)-(PWA), NNLO$_\text{sat}$, and $\Delta$NNLO$_\text{GO}$(450) refer to the chiral interactions considered in Ref.~\cite{Payne:2019wvy}.}
\label{fig:Fweak_Ar}
\end{figure}

\begin{figure}[t]  
	\begin{center}
	\includegraphics[width=0.48\textwidth,clip]{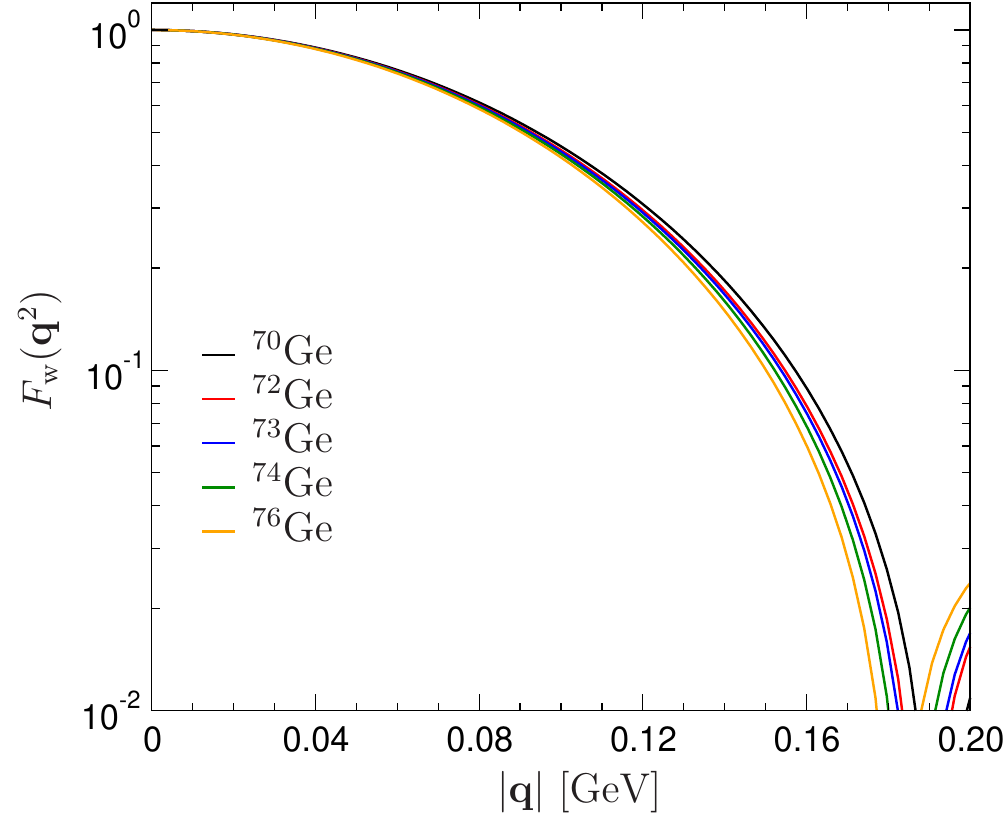}
	\end{center}
\caption{Shell-model results for the weak form factor of germanium.}
\label{fig:Fweak_Ge}
\end{figure}

\begin{figure}[t]
	\begin{center}
	\includegraphics[width=0.48\textwidth,clip]{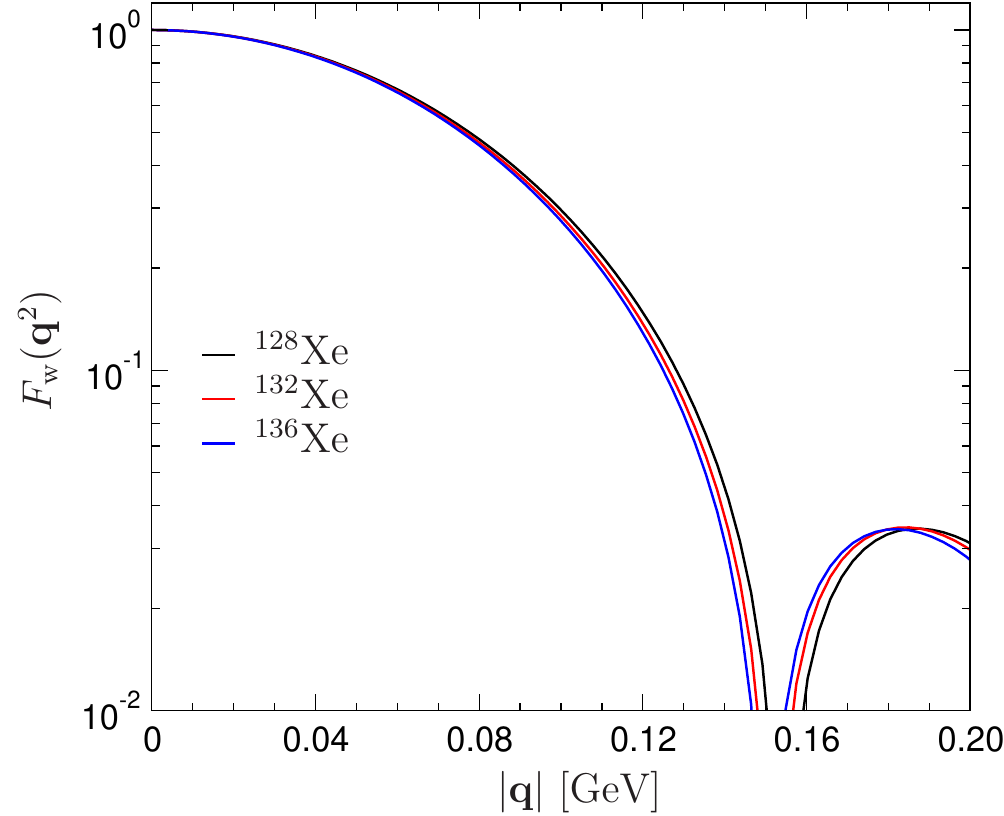}
	\end{center}
\caption{Shell-model results for the weak form factor of xenon (we only show selected isotopes for better visibility).}
\label{fig:Fweak_Xe}
\end{figure}

The corresponding results for the weak form factors are shown in Figs.~\ref{fig:Fweak_CsI}--\ref{fig:Fweak_Xe}. In each case, we show the shell-model results for the modulus of the weak form factor including all corrections given in Eq.~\eqref{Fw_definition}.
Coherence is kept until larger momentum transfers in lighter nuclei with smaller neutron radius, see Fig.~\ref{fig:Fweak_CsI}.
For germanium and xenon isotopes, Figs.~\ref{fig:Fweak_Ge} and \ref{fig:Fweak_Xe} show the difference between the weak form factors of stable isotopes.

In the case of $^{40}$Ar, Fig.~\ref{fig:Fweak_Ar} compares our results to the RMF calculation of Ref.~\cite{Yang:2019pbx} as well as ab initio results from coupled-cluster theory~\cite{Payne:2019wvy}. 
All calculated weak form factors give similar results, within the uncertainty band estimated in Ref.~\cite{Payne:2019wvy}.
This suggests that uncertainties in the neutron distribution are relatively small, in contrast to the assumptions in Ref.~\cite{AristizabalSierra:2019zmy}.
We stress that apart from the nuclear structure, minor differences in the weak form factor arise from the precise input for the hadronic matrix elements and weak charges, primarily the proton charge radius, for which Refs.~\cite{Yang:2019pbx,Payne:2019wvy} use $\langle r_E^2\rangle^p\simeq 0.77\fm^2$. 

\subsection{Neutrino scattering}
\label{sec:CEvNS}

The dominant contribution to the CE$\nu$NS cross section in the SM involves the same nuclear form factor as in the case of PVES, since apart from overall prefactors the combination of Wilson coefficients, hadronic matrix elements, and nuclear structure factors remains unchanged. This dominant piece of the differential cross section takes the form
\beq
\label{CEvNS_coherent}
\frac{\diff \sigma_A}{\diff T}\bigg|_\text{coherent}=\frac{G_F^2\mA}{4\pi}\bigg(1-\frac{\mA T}{2E_\nu^2}\bigg)Q_\text{w}^2\big|F_\text{w}(\qq^2)\big|^2,
\eeq
where $E_\nu$ is the energy of the incoming neutrino and the nuclear recoil
\beq
\label{def_T}
T=E_\nu-E_\nu'=-\frac{t}{2\mA}
\eeq
takes values in $[0,2E_\nu^2/(\mA+2E_\nu)]$. Terms of order $T/E_\nu\lesssim 2E_\nu/\mA$ are usually neglected due to typical neutrino energies $E_\nu\lesssim 50\MeV$. 
The cross section in Eq.~\eqref{CEvNS_coherent} represents the truly ``coherent'' contribution, in the sense that the nuclear structure factors that enter the definition of $F_\text{w}$, see Eq.~\eqref{Fw_definition}, indeed scale with $Z$ and $N$ ($\F^M$) or at least can receive some partial coherent enhancement with respect to closed shells ($\F^{\Phi''}$). Two-body corrections to Eq.~\eqref{CEvNS_coherent} again only arise at the loop level, and are thus significantly suppressed in the chiral expansion. 

Before extending Eq.~\eqref{CEvNS_coherent} to the axial-vector responses, we comment on some details of the derivation as well as subleading kinematic effects. The starting point is the leptonic trace
\begin{align}
\label{leptonic}
L^{\mu\nu}&=\Tr\big(\slashed{k'}\gamma^\mu P_L\slashed{k}\gamma^\nu P_L\big)\notag\\
&=2\big(k^\mu k'^\nu+k'^\mu k^\nu-g^{\mu\nu}k\cdot k'+i\eps^{\mu\nu\alpha\beta}k_\alpha k'_\beta\big),
\end{align}
whose components determine the spin sums
\begin{align}
\label{spin_sums}
 \sum_\text{spins}l_0l_0^*&=L_{00}=2E_\nu^2\bigg(2-\frac{\mA T}{E_\nu^2}-\frac{2T}{E_\nu}\bigg)+\Order\big(T^2\big),\notag\\
 \sum_\text{spins}l_3l_3^*&=L_{33}=2E_\nu^2\frac{T}{\mA}+\Order\big(T^2\big),\notag\\
 \sum_\text{spins}l_0l_3^*&=L_{03}=2E_\nu^2\sqrt{\frac{2T}{\mA}}+\Order\big(T^{3/2}\big),\\
  \sum_\text{spins} \llvec\cdot \llvec^*&=L_{ii}=2E_\nu^2\bigg(2+\frac{\mA T}{E_\nu^2}-\frac{2T}{E_\nu}\bigg)+\Order\big(T^2\big),\notag\\
 \sum_\text{spins} (\llvec\times \llvec^*)_3&=\eps_{3ij}L_{ij}=-4iE_\nu\sqrt{2\mA T}+\Order\big(T^{3/2}\big).\notag
\end{align}
The spherical components are defined with respect to the direction of $\qq=\kk'-\kk$, e.g.,
\begin{align}
k_3&=\frac{\kk\cdot\qq}{|\qq|}=-\frac{T(\mA+E_\nu)}{\sqrt{T(2\mA+T)}},\notag\\
k_3'&=\frac{\kk'\cdot\qq}{|\qq|}=\frac{T(\mA+T-E_\nu)}{\sqrt{T(2\mA+T)}}.
\end{align}
In particular, the combination $L_{33}$ is strongly suppressed by $T/\mA\lesssim 2E_\nu^2/\mA^2$, while $L_{03}$ or the additional terms in $L_{00}$ and $L_{ii}$ are only suppressed by $T/E_\nu\lesssim 2E_\nu/\mA$.
In consequence, the longitudinal multipoles in Eq.~\eqref{multipole_expansion} can be safely neglected.
The interference with the Coulomb multipoles could in principle become relevant,
but the longitudinal multipoles involve an additional suppression by $q^0/|\qq|=-\sqrt{T/(2\mA)}\lesssim - E_\nu/\mA$ from the application of current conservation, see Eq.~\eqref{multipoles_matching}. 
Accordingly, all potentially relevant subleading kinematic effects can be taken into account by  
\beq
\bigg(1-\frac{\mA T}{2E_\nu^2}\bigg)\to
\bigg(1-\frac{\mA T}{2E_\nu^2}-\frac{T}{E_\nu}
\bigg)
\eeq
in Eq.~\eqref{CEvNS_coherent}.

\begin{table}[t]
	\centering
	\renewcommand{\arraystretch}{1.3}
	\begin{tabular}{c|rrrr}
		\toprule
		& $^{19}$F & $^{23}$Na & $^{73}$Ge & $^{127}$I\\
$\langle \spin_p\rangle$ &  $0.478$ & $0.224$ & $0.032$ & $0.346$ \\ 
$\langle \spin_n\rangle$ & $-0.002$ & $0.024$ & $0.439$ & $0.031$ \\\colrule
& $^{129}$Xe &  $^{131}$Xe &  $^{133}$Cs &\\
$\langle \spin_p\rangle$ & $0.010$ & $-0.009$ & $-0.343$ & \\ 
$\langle \spin_n\rangle$ & $0.329$ & $-0.272$ & $0.001$ &\\
		\botrule
	\end{tabular}
	\renewcommand{\arraystretch}{1.0}
	\caption{Shell-model proton ($\langle \spin_p\rangle$) and neutron ($\langle \spin_n\rangle$) spin expectation values for the odd-mass isotopes considered in this work.
	}
	\label{tab:spin}
\end{table}

Next, there could be interference terms between the vector and axial-vector responses.  The vector contributions to the transverse multipoles vanish for $T\to 0$ and are not coherent, so the only potentially relevant interferences arise from the longitudinal and Coulomb multipoles. However, all such interferences vanish due to Eq.~\eqref{multipoles_matching}.
 
Therefore, the dominant correction to
Eq.~\eqref{CEvNS_coherent} comes solely from the axial-vector part of the interaction. This contribution becomes relevant for precision studies of nuclei with nonvanishing spin, especially, because in contrast to other less relevant corrections their contribution remains finite in the limit $T\to 0$. 
The SD structure factors are obtained by adapting the formalism from Ref.~\cite{Klos:2013rwa}, most notably, by only keeping the transverse electric multipoles, due to the strong suppression of the longitudinal ones 
(transverse magnetic multipoles do not contribute to elastic scattering due to time reversal). Collecting the kinematic factors, the resulting contribution to the CE$\nu$NS cross section takes the form
\begin{align}
\frac{\diff \sigma_A}{\diff T}\bigg|_\text{SD}&=\frac{2\mA}{2J+1}\bigg(2+\frac{\mA T}{E_\nu^2}-\frac{2T}{E_\nu}\bigg)\\
&\hspace{-12pt}\times\Big(\big(g_A^0\big)^2 S_{00}^\mathcal{T}(\qq^2)+g_A^0 g_A^1 S_{01}^\mathcal{T}(\qq^2)+ \big(g_A^1\big)^2 S_{11}^\mathcal{T}(\qq^2)\Big),\notag
\end{align}
where the structure factors $S_{ij}^\mathcal{T}(\qq^2)$ are the same as for dark matter except that longitudinal multipoles need to be omitted, see Sec.~\ref{sec:SD_structure factors} as well as Apps.~\ref{app:multipole} and \ref{app:nuc_responses} for the precise definitions. In particular, the normalizations are related to $\langle \spin_N\rangle$, the nucleon (proton and neutron) spin expectation values:\footnote{Note that Eq.~\eqref{Fsigmanorm} includes an additional factor $1/2$ compared to the $\qq=0$ limit of the standard definitions of the $\Sigma'$, $\Sigma''$ operators in Eqs.~\eqref{Sigma'}--\eqref{Sigma''} (the same is true for the full $\mathcal{F}_N^{\Sigma'_L}(\qq^2)$, $\mathcal{F}_N^{\Sigma''_L}(\qq^2)$). This factor is compensated by the factor 2 in Eqs.~\eqref{Sp}--\eqref{Sn}, which is needed for consistency with the definition of $S_{ij}$ in Eqs.~\eqref{Sij}--\eqref{S01} in the literature.}
\beq
\label{Fsigmanorm}
\mathcal{F}_N^{\Sigma'_1}(0)=\sqrt{2}\, \mathcal{F}_N^{\Sigma''_1}(0)=\sqrt{\frac{2}{3}}\sqrt{\frac{(2J+1)(J+1)}{4\pi J}} \langle \spin_N\rangle.
\eeq
We have obtained the nuclear responses $\mathcal{F}_N^{\Sigma'_1}(\qq)$ and  $\mathcal{F}_N^{\Sigma''_1}(\qq)$
and the corresponding spin expectation values with the nuclear shell model calculations described in Sec.~\ref{sec:PVES}.
The results for the spin expectation values are given in Table~\ref{tab:spin}, see also Refs.~\cite{Menendez:2012tm,Klos:2013rwa}. 
The isoscalar/isovector coefficients are
\beq
\label{def_axial_vector}
g_A^0=\frac{g_A^p+g_A^n}{2},\qquad g_A^1=\frac{g_A^p-g_A^n}{2},
\eeq
where $g_A^N=\sum_q C_q^A g_A^{q,N}$. In the SM we have, using Eqs.~\eqref{Wilson_SM}, \eqref{gAiso}, and \eqref{gA},
\begin{align}
g_A^p&=\frac{G_F}{\sqrt{2}}\big(g_A - g_A^{s,N}\big),& 
g_A^n&=-\frac{G_F}{\sqrt{2}}\big(g_A + g_A^{s,N}\big),\notag\\
g_A^0&=-\frac{G_F}{\sqrt{2}}g_A^{s,N},& 
g_A^1&=\frac{G_F}{\sqrt{2}}g_A,
\end{align}
so that the full expression for the cross section becomes
\begin{align}
\label{CEvNS_SM}
 \frac{\diff \sigma_A}{\diff T}&=\frac{G_F^2\mA}{4\pi}\bigg(1-\frac{\mA T}{2E_\nu^2}-\frac{T}{E_\nu}\bigg)Q_\text{w}^2\big|F_\text{w}(\qq^2)\big|^2\notag\\
 &+\frac{G_F^2\mA}{4\pi}\bigg(1+\frac{\mA T}{2E_\nu^2}-\frac{T}{E_\nu}\bigg)F_A(\qq^2),
 %&+\frac{G_F^2\mA}{2J+1}\bigg(2+\frac{\mA T}{E_\nu^2}-\frac{2T}{E_\nu}\bigg)\notag\\
 %&\hspace{-12pt}\times\Big(\big(g_A^{s,N}\big)^2 S_{00}^\mathcal{T}(\qq^2)-g_A g_A^{s,N} S_{01}^\mathcal{T}(\qq^2)+ \big(g_A\big)^2 S_{11}^\mathcal{T}(\qq^2)\Big),\notag
\end{align}
where
\begin{align}
&F_A(\qq^2)=\frac{8\pi}{2J+1}\\
&\hspace{11pt}\times\Big(\big(g_A^{s,N}\big)^2 S_{00}^\mathcal{T}(\qq^2)-g_A g_A^{s,N} S_{01}^\mathcal{T}(\qq^2)+ \big(g_A\big)^2 S_{11}^\mathcal{T}(\qq^2)\Big)\notag
\end{align}
is the axial-vector analog of $\big|F_\text{w}(\qq^2)\big|^2$. As expected, 
the dominant SD correction arises from the isovector component, with the normalization
\beq
F_A(0)=\frac{4}{3}g_A^2\frac{J+1}{J}\big(\langle \spin_p\rangle-\langle \spin_n\rangle\big)^2,
\eeq
when strangeness and two-body corrections are neglected. 
The induced pseudoscalar form factor $G_P(t)$ only contributes to the longitudinal multipoles, see Eq.~\eqref{multipoles_matching}. Since $g_A$ factorizes, the radius corrections from Eq.~\eqref{axial_radius} are usually absorbed into the structure factors, as are corrections from two-body currents, to which we will turn in the next subsection.

\subsection{Improved treatment of axial-vector two-body currents}
\label{sec:axial_vector_responses}

Axial-vector currents are responsible for SD scattering.
In the nonrelativistic limit the leading one-body (1b) currents are given by
\begin{equation}
\label{axial1bc}
\JJ^3_{i,{\rm 1b}} = \frac{1}{2} \, \tau_i^3 \biggl( G_A^3(\qq^2) \, \sig_i
- \frac{G_P^3(\qq^2)}{4 \mN^2} \,
(\qq \cdot \sig_{i}) \, \qq \biggl) ,
\end{equation}
so that axial responses are driven by the nucleon spin ${\bf S}_i=\sig_i/2$,
as indicated by Table~\ref{tab:nuclear_structure_factors}.

A sizable correction to the leading one-body terms comes from subleading axial-vector two-body currents~\cite{Hoferichter:2015ipa}.
In medium-mass and heavy nuclei,
these contributions have been evaluated in previous studies of $\beta$ decays~\cite{Menendez:2011qq,Gysbers:2019uyb} and WIMP--nucleus scattering~\cite{Menendez:2012tm,Klos:2013rwa}.
However,
%none of these previous works has considered all axial-vector two-body contributions up to third order in chiral EFT as presented in Ref.~\cite{Hoferichter:2015ipa}.
%In particular, 
the studies of SD WIMP scattering off nuclei focus on
pion-exchange two-body currents proportional to the low-energy couplings $c_3$, $c_4$, and $c_6$~\cite{Menendez:2012tm,Klos:2013rwa} and neglected the contact two-body axial-vector current proportional to the couplings $d_1$, $d_2$ ~\cite{Hoferichter:2015ipa},
which is only included in the $|\qq|=0$ limit in $\beta$ decay~\cite{Menendez:2011qq,Gysbers:2019uyb}.
%All these works also neglect pion-pole contributions proportional
%to the couplings $c_i$ and $d_i$.

Here we improve previous studies by including all pion-exchange, pion-pole, and contact terms derived in Ref.~\cite{Hoferichter:2015ipa}:
\begin{align}
\label{axial2bc}
\JJ^3_{12} &= -\frac{g_A}{2F^2_\pi} \, [\tau_1\times\tau_2]^3 \nonumber \\
&\quad
\biggl[ 
c_4\Bigl(1-\frac{\qq}{\qq^2+M_\pi^2}{\qq\cdot}\Bigr)
(\sig_1\times\kk_2)
\nonumber \\
&\quad +\frac{c_6}{4}
(\sig_1\times\qq) 
+i \, \frac{{\pp_1+\bf p}'_1}{4\mN}
 \biggr]
\frac{\sig_2\cdot\kk_2}{M_\pi^2+k_2^2}  \nonumber \\
&-\frac{g_A}{F^2_\pi} \,\tau_2^3
\biggl[c_3 \Bigl(1-\frac{\qq}{\qq^2+M_\pi^2}{\qq\cdot}\Bigr) \,\kk_2
\nonumber \\
&\quad +2c_1M_\pi^2 \frac{\qq}{\qq^2+M_\pi^2} \biggr]  \, \frac{
	\sig_2\cdot\kk_2}{M_\pi^2+k_2^2}  \nonumber \\
&- d_1 \,\tau_1^3  \Bigl(1-\frac{\qq}{\qq^2+M_\pi^2}{\qq\cdot}\Bigr)
\sig_1 +(1\longleftrightarrow2) \nonumber \\
&- d_2 (\tau_1\times\tau_2)^3 (\sig_1 \times \sig_2)
\Bigl(1-{\cdot \qq}\frac{\qq}{\qq^2+M_\pi^2}\Bigr),
\end{align}
where $\kk_i=\pp'_i-\pp_i$, $\qq = -\kk_1-\kk_2$, and $(1\longleftrightarrow2)$ applies to the entire expression except for the last line.
Relativistic $1/\mN$ corrections to Eq.~\eqref{axial2bc}, besides the term proportional to $\pp_1+\pp_1'$, can be absorbed into $c_4\to c_4+1/(4\mN)$, $c_6\to c_6+1/\mN$, where we use a dimensionful 
$c_6$ for consistency with the previous literature on the axial current (note that our choice of $c_6$ corresponds to $c_6/\mN$ in the conventions of Ref.~\cite{Bernard:1992qa}). 
In the counting of Refs.~\cite{Hoferichter:2015ipa,Krebs:2016rqz} these relativistic corrections are formally of higher order, but we keep them both for consistency with Ref.~\cite{Gysbers:2019uyb} and in analogy to our treatment of higher-order effects in the $c_i$, see below. 

Following Refs.~\cite{Menendez:2012tm,Klos:2013rwa} we approximate the two-nucleon currents by a normal-ordering approximation with respect to spin-isospin symmetric reference state with density $\rho=2k_{\rm F}^3/(3\pi^2)$
($k_{\rm F}$ is the Fermi momentum)
\begin{align}
\label{normalordering}
\JJ^{\rm eff}_{i,{\rm 2b}} =\sum_j(1-P_{ij}) \JJ^3_{ij},
\end{align}
where $P_{ij}$ is the exchange operator and the sum is performed over the second nucleon $j$.

As a result, axial-vector two-body currents transform into effective one-body currents~\cite{Klos:2013rwa,Klos:2014}:
\begin{align}
&\JJ^{\text{eff},\sigma}_{i,\text{2b}}(\rho,\qq,\PP)
=-g_A \sig_i \, \frac{\tau^3_i}{2} \, \frac{\rho}{2 F^2_\pi}
\nonumber \\
&\quad \times \Biggl( -\frac{1}{3}
\Bigl(
c_3-\frac{1}{4\mN}\Bigr)
\Bigr[ I_1^{\sigma}(\rho,|\PP-\qq|)+I_1^{\sigma}(\rho,|\PP
+\qq|) \Bigr] \nonumber \\
&\qquad+\frac{c_4}{3} 
 \Big[
3I_2^{\sigma}(\rho,|\PP-\qq|)-I_1^{\sigma}(\rho,|\PP-\qq|)
\nonumber \\
&\qquad+3I_2^{\sigma}(\rho,|\PP+\qq|)-I_1^{\sigma}(\rho,|\PP+
\qq|) \Big] \nonumber \\
&\qquad+ \frac{c_6}{12}
 \biggl[I_{c6}(\rho,
|\PP-\qq|) \, \frac{\qq\cdot(\PP-\qq)}{
	(\PP-\qq)^2} \nonumber \\
&\qquad-I_{c6}(\rho,|\PP+\qq|) \, \frac{\qq\cdot(\PP
	+\qq)}{(\PP+\qq)^2} \biggr]-\frac{c_D}{2g_A\Lambda_\chi}
\Biggl) ,
\label{scurrent}\\
&\JJ^{\text{eff},P}_{i,\text{2b}}(\rho,\qq,\PP) = 
-g_A \, \frac{\tau^3_i}{2} \, (\qq\cdot\sig_i) \, \qq 
\, \frac{\rho}{2 F_\pi^2} \nonumber \\
&\times \Biggl(
4\bigl(c_3-2c_1\bigr)\,\frac{M_\pi^2 }{(M_\pi^2+\qq^2)^2} \nonumber \\
&\quad - \frac{1}{3} \Bigl(c_3+c_4-\frac{1}{4\mN}\Bigr) \frac{I^P(\rho,|\PP-\qq|)
	+I^P(\rho,|\PP+\qq|)}{\qq^2} \nonumber \\
&\quad+ \frac{1}{3} \bigl(c_3+c_4\bigr) \frac{1}{M_\pi^2+\qq^2} \nonumber \\
&\qquad \times\Bigr[ I_1^{\sigma}(\rho,|\PP-\qq|)+I_1^{\sigma}(\rho,|\PP
+\qq|) 
\nonumber \\
&\qquad \quad  
+\frac{\qq^2I^{P}(\rho,|\PP-\qq|)}{(\PP-\qq)^2}
+\frac{\qq^2I^{P}(\rho,|\PP+\qq|)}{(\PP+\qq)^2}
\biggr] \nonumber \\
&\quad-c_4\frac{1}{M_\pi^2+\qq^2} \Bigr[ I_2^{\sigma}(\rho,|\PP-\qq|)+I_2^{\sigma}(\rho,|\PP
+\qq|) \Bigr] \nonumber \\
&\quad + \Bigl(\frac{c_6}{12}
-\frac{2}{3}\frac{c_1M_\pi^2}{M_\pi^2+\qq^2} \Bigr) \nonumber \\
&\qquad \times\biggl[
\frac{I_{c6}(\rho,|\PP-\qq|)}{(\PP-\qq)^2}
+\frac{I_{c6}(\rho,|\PP+\qq|)}{(\PP+\qq)^2}
\biggr] \nonumber \\
&\quad+\frac{c_D}{2g_A\Lambda_\chi} \frac{1}{M_\pi^2+\qq^2}
\Biggr) .\label{Pcurrent}
\end{align}
These two effective currents have the same structure as the two terms in the leading one-body current, Eq.~\eqref{axial1bc}, so they can be treated in the same way.

The currents in Eqs.~\eqref{scurrent}--\eqref{Pcurrent} depend on the nuclear density $\rho$, the momentum transfer $\qq$,
and the combined momentum $\PP$.
Because the dependence on $\PP$ is small~\cite{Klos:2013rwa},
in practice we evaluate the expressions taking $\PP=0$.
Likewise, we neglect additional effective one-body currents proportional
to $\PP$ and $\PP\cdot\sig_i$.
The functions $I^\sigma_1(\rho,K)$,
$I^\sigma_2(\rho,K)$, $I^P(\rho,K)$, and $I_{c6}(\rho,K)$
appear due to the summation over occupied states in the exchange terms in Eq.~\eqref{normalordering}.
They can be expressed as integrals, with analytical expressions given in Ref.~\cite{Klos:2013rwa}.

In the $\PP=0$ approximation, the combined effective currents can be written in analogy to Eq.~\eqref{axial1bc}
\begin{align}
\label{eff2bcombined}
&\JJ^{\text{eff}}_{i,\text{2b}}(\rho,\qq)
=g_A \, \frac{\tau^3_i}{2} \bigg[ \delta a(\qq^2)\, \sig_i + \frac{\delta a^P(\qq^2)}{\qq^2} (\qq\cdot\sig_i)\qq \bigg],
\end{align}
where
\begin{align}
\label{deltaa}
\delta a(\qq^2) &= -\frac{\rho}{F^2_\pi} \, 
\biggl[ \frac{c_4}{3}
\Bigl[3I_2^{\sigma}(\rho,|\qq|)-I_1^{\sigma}(\rho,|\qq|) \Bigr] \nonumber \\
&-\frac{1}{3}\Bigl(c_3-\frac{1}{4\mN}\Bigr) 
I_1^{\sigma}(\rho,|\qq|) -  \frac{c_6}{12} I_{c6}(\rho,|\qq|) \notag\\
&-\frac{c_D}{4g_A\Lambda_\chi} \biggl], \\
\delta a^P(\qq^2)&=
\frac{\rho}{F^2_\pi} \biggl[ 
-2\bigl(c_3-2c_1\bigr)\,\frac{M_\pi^2\,\qq^2 }{(M_\pi^2+\qq^2)^2} \nonumber \\
&+\frac{1}{3} \Bigl(c_3+c_4-\frac{1}{4\mN}\Bigr)\, I^P(\rho,|\qq|) 
\nonumber \\
&- \Bigl(\frac{c_6}{12}
-\frac{2}{3}\frac{c_1M_\pi^2}{M_\pi^2+\qq^2} \Bigr) \, I_{c6}(\rho,|\qq|)  \nonumber \\
& -\frac{\qq^2}{M_\pi^2+\qq^2} \Bigr[
\frac{c_3}{3} \bigr[ I_1^{\sigma}(\rho,|\qq|) +I^{P}(\rho,|\qq|)\bigr] \nonumber \\
&\qquad +\frac{c_4}{3}\bigr[ I_1^{\sigma}(\rho,|\qq|) +I^{P}(\rho,|\qq|)-3I_2^{\sigma}(\rho,|\qq|)\bigr]
\Bigr]    \nonumber \\
&-\frac{c_D}{4g_A\Lambda_\chi}\frac{\qq^2}{M_\pi^2+\qq^2}
\biggr].
\label{deltaap}
\end{align}

For $\beta$ decays $\qq\simeq0$,
and axial-vector two-body currents have been studied 
beyond the normal-ordering approximation in Eq.~\eqref{normalordering}~\cite{Gysbers:2019uyb}.
The approximation for $\JJ^{\rm eff}_{i,{\rm 2b}}$ was found to be very good when taking $\rho\sim0.10~{\rm fm}^{-3}$, which is a typical value for the density of the nuclear surface. 
Based on this, for our evaluation of the nuclear structure factors we consider the density range $\rho=0.09\ldots 0.11 \, {\rm fm}^{-3}$.
This range includes slightly lower densities, but is consistent with the one considered in Refs.~\cite{Menendez:2012tm,Klos:2013rwa}.

The contributions from two-body currents in Eqs.~\eqref{scurrent}--\eqref{Pcurrent}
depend on the low-energy couplings
$c_1$, $c_3$, $c_4$, $c_6$, and $c_D$.
Due to antisymmetrization of the currents, the two couplings of the contact two-body term combine into a single contribution proportional to $c_D=-4(d_1+2d_2)/(F_\pi^2\,\Lambda_\chi)$.
The values of $c_i$, $c_D$ to be used should in principle be given by the nuclear interaction used to solve the many-body problem for the nucleus of interest.
However, accurate many-body calculations using chiral interactions that depend explicitly
on $c_i$, $c_D$ are still not available for all nuclei discussed in this work.
Instead, our results are based on many-body calculations that use shell-model Hamiltonians, which, despite being based on nucleon--nucleon interactions, are modified by phenomenological adjustments in order to improve their description of the nuclear structure of selected regions of nuclei.
Therefore, we cannot use consistent $c_i$, $c_D$ couplings between the nuclear interactions and the two-nucleon currents given in Eqs.~\eqref{scurrent}--\eqref{Pcurrent}.

\begin{table}[t]
	\renewcommand{\arraystretch}{1.3}
	\begin{center}
		\begin{tabular}{l r}\toprule
			$c_1$ [GeV$^{-1}$] & $-1.20(17)$\\
			$c_3$ [GeV$^{-1}$] & $-4.45(86)$\\
			$c_4$ [GeV$^{-1}$] & $2.96(70)$\\
			$c_6$ [GeV$^{-1}$] & $5.01(1.06)$\\\colrule
			$c_D$  & $-6.08\ldots0.30$\\\colrule
			$\rho$ [fm$^{-3}$] & $0.09\ldots0.11$\\\botrule
		\end{tabular}
		\caption{Nuclear density $\rho$ and low-energy couplings $c_i$ and $c_D$ used in this work. The smallest (largest) value of $c_D$ is only reached for the lowest (highest) density $\rho=0.09$~fm$^{-3}$ ($\rho=0.11$~fm$^{-3}$) and 30\% (20\%) contribution of two-body axial currents at $|\qq|=0$. The values for $c_{1,3,4,6}$ include the leading-loop effects and relativistic corrections as described in the main text. The chiral scale in the definition of $c_D$ is set to $\Lambda_\chi=700\MeV$.
			\label{ci_values}}
	\end{center}
\end{table}

Our strategy is as follows.
First, we use the values for $c_1$, $c_3$, and $c_4$ determined in the Roy--Steiner-equation analysis of $\pi N$ scattering~\cite{Hoferichter:2015tha,Hoferichter:2015hva}. This improved determination of the $c_i$ values allows us to obtain results with reduced theoretical uncertainties compared to Refs.~\cite{Menendez:2012tm,Klos:2013rwa}, which considered a broad range of $c_3$ and $c_4$
(the smaller $c_1$ contributions are included for the first time in this work). In fact, at a given chiral order the uncertainties in the $c_i$ are now negligible, with the main uncertainty arising from the chiral expansion. Strictly speaking, one should use the next-to-leading-order values from Refs.~\cite{Hoferichter:2015tha,Hoferichter:2015hva} to be consistent with the chiral order we use for the axial-vector current, but this assumes that the latter is affected by large loop corrections in the same way as $\pi N$ scattering, which is known not to be the case. Instead, we make use of the fact that the two-nucleon axial-vector current is matched to the three-nucleon force~\cite{Krebs:2016rqz}, in such a way that the
leading loop corrections in the axial-vector current coincide with the ones in the three-nucleon force~\cite{Bernard:2007sp,Ishikawa:2007zz}. These corrections can be represented by a simple shift $\delta c_i$~\cite{Epelbaum:2008ga}
\beq
\delta c_1=-\frac{\gA^2\mpi}{64\pi\Fpi^2},\qquad 
\delta c_3=-\delta c_4=\frac{\gA^4\mpi}{16\pi\Fpi^2}.
\eeq
The values shown in Table~\ref{ci_values} are then obtained as the combination of the next-to-next-to-leading-order values from Refs.~\cite{Hoferichter:2015tha,Hoferichter:2015hva} in combination with these $\delta c_i$ (as well as the relativistic correction for $c_4$), and the uncertainties represent the shifts between the two chiral orders. 
The value of $c_6$ is related to the isovector magnetic moments via~\cite{Bernard:1992qa}
\beq
\label{c6_matching}
c_6=\frac{\kappa^p-\kappa^n}{\mN}+\frac{g_A^2\mpi}{4\pi\Fpi^2},
\eeq
where we have indicated the leading loop correction. Similarly to the other $c_i$, this correction is large despite being formally of higher order (in part due to the enhancement by a factor of $\pi$~\cite{Baru:2012iv}). However, similar corrections arise from chiral loops in the axial-vector current~\cite{Krebs:2016rqz,Baroni:2015uza}, the dominant of which can again be represented as a shift in $c_6$ 
\beq
\delta c_6=-\frac{g_A^2\mpi}{4\pi\Fpi^2}
\eeq
and cancels the matching correction in Eq.~\eqref{c6_matching}. Including the relativistic corrections discussed before, we will thus equate $c_6=(\kappa^p-\kappa^n+1)/\mN=5.01$, as given in Table~\ref{ci_values}.

We then fix the value of the contact coupling $c_D$, while at the same time correcting for the shortcomings of our phenomenological calculations.
Shell-model nuclear matrix elements involving the axial-vector current typically overestimate experiment~\cite{MartinezPinedo:1996vz} by about 20\% to 30\%.
Recently, Ref.~\cite{Gysbers:2019uyb} showed for $\beta$ decay (where it is sufficient to take $|\qq|=0$) that this is because of a combination of missing two-body axial-vector currents, see
Eq.~\eqref{axial2bc}, and additional nuclear correlations that are beyond the standard shell-model approach.
In order to account for this, we adjust the value of $c_D$ so that our shell-model calculations receive a contribution from two-nucleon currents such that, at $|\qq|=0$, 
Eq.~\eqref{scurrent} reduces the leading term in Eq.~\eqref{axial1bc} in the range 20\% to 30\%.
The $\qq$ dependence of the effective two-body currents is the one predicted by Eqs.~\eqref{scurrent}--\eqref{Pcurrent}.
Since the leading contribution from two-body axial-vector currents comes from
the pion-exchange part proportional to $c_3$ and $c_4$,
the part considered in Refs.~\cite{Menendez:2012tm,Klos:2013rwa},
our results are consistent with these previous calculations.

The values of $c_i$ and $c_D$ used in this work are summarized in Table~\ref{ci_values},
where the extreme values $c_D=-6.08$ ($c_D=0.30$) only correspond to the low density $\rho=0.09$~fm$^{-3}$ (high density $\rho=0.11$~fm$^{-3}$).
In practice, we neglect the remaining uncertainties in the $c_i$ due to effects from higher chiral orders not captured here, as those are subleading compared to 
the uncertainty in the range of $c_D$ values,
which also depend on the nuclear density $\rho$.
Ultimately, our uncertainty depends on the range imposed on the impact of the two-body currents at $|\qq|=0$, $20\%$--$30\%$, as estimated from $\beta$ decay~\cite{MartinezPinedo:1996vz,Gysbers:2019uyb}.

\subsection{Spin-dependent responses for CE$\boldsymbol{\nu}$NS and dark matter}
\label{sec:SD_structure factors}

The nuclear responses for CE$\nu$NS and SD dark matter scattering off nuclei can be expressed in terms of the transverse and longitudinal SD structure factors $\mathcal{F}_\pm^{\Sigma'_L}(\qq^2)$ and $\mathcal{F}_\pm^{\Sigma''_L}(\qq^2)$, respectively.
For CE$\nu$NS, only the transverse component contributes, while for dark matter scattering both longitudinal and transverse parts need to be taken into account.

The expressions are given by
\begin{align}
\label{Sij}
S_{00} &= S_{00}^{\mathcal{T}} + S_{00}^{\mathcal{L}}\notag\\
&= \sum_L \Big[\mathcal{F}_+^{\Sigma'_L}(\qq^2)\Big]^2 + \sum_L \Big[\mathcal{F}_+^{\Sigma''_L}(\qq^2)\Big]^2, \\
S_{11} &= S_{11}^{\mathcal{T}} + S_{11}^{\mathcal{L}} \nonumber \\
&= \sum_L \Big[[1+\delta'(\qq^2)]\mathcal{F}_-^{\Sigma'_L}(\qq^2)\Big]^2 \notag\\
&+ \sum_L \Big[[1+\delta''(\qq^2)]\mathcal{F}_-^{\Sigma''_L}(\qq^2)\Big]^2, \\
S_{01} &= S_{01}^{\mathcal{T}} + S_{01}^{\mathcal{L}} \nonumber \\
&= \sum_L 2[1+\delta'(\qq^2)]\mathcal{F}_+^{\Sigma'_L}(\qq^2)\,\mathcal{F}_-^{\Sigma'_L}(\qq^2)
\nonumber \\ 
&+ \sum_L 2[1+\delta''(\qq^2)]\mathcal{F}_+^{\Sigma''_L}(\qq^2)\,\mathcal{F}_-^{\Sigma''_L}(\qq^2), \label{S01}
\end{align}
which can be expressed in terms of the proton/neutron instead of the isoscalar/isovector basis as
\begin{align}
S_p & = S_{p}^{\mathcal{T}} + S_{p}^{\mathcal{L}} \label{Sp}\\
&= \sum_L \Big[2\mathcal{F}_p^{\Sigma'_L}(\qq^2)+\delta'(\qq^2)\Big(\mathcal{F}_p^{\Sigma'_L}(\qq^2)-\mathcal{F}_n^{\Sigma'_L}(\qq^2)\Big) \Big]^2 \nonumber \\ 
&+ \sum_L \Big[2\mathcal{F}_p^{\Sigma''_L}(\qq^2)+\delta''(\qq^2)\Big(\mathcal{F}_p^{\Sigma''_L}(\qq^2)-\mathcal{F}_n^{\Sigma''_L}(\qq^2)\Big) \Big]^2, \notag\\
S_n & = S_{n}^{\mathcal{T}} + S_{n}^{\mathcal{L}}  \label{Sn}\\
&= \sum_L \Big[2\mathcal{F}_n^{\Sigma'_L}(\qq^2)-\delta'(\qq^2)\Big(\mathcal{F}_p^{\Sigma'_L}(\qq^2)-\mathcal{F}_n^{\Sigma'_L}(\qq^2)\Big) \Big]^2 \nonumber \\ 
&+ \sum_L \Big[2\mathcal{F}_n^{\Sigma''_L}(\qq^2)-\delta''(\qq^2)\Big(\mathcal{F}_p^{\Sigma''_L}(\qq^2)-\mathcal{F}_n^{\Sigma''_L}(\qq^2)\Big) \Big]^2,\notag
\end{align}
where the proton/neutron combinations are related to the isospin ones analogously to Eq.~\eqref{Fpn}
\begin{align}
\label{Fpnsigma}
\F^{\Sigma'_L}_\pm(\qq^2)&=\F^{\Sigma'_L}_p(\qq^2)\pm \F^{\Sigma'_L}_n(\qq^2),\notag\\
\F^{\Sigma''_L}_\pm(\qq^2)&=\F^{\Sigma''_L}_p(\qq^2)\pm \F^{\Sigma''_L}_n(\qq^2).
\end{align}

The terms $\delta'(\qq^2),\delta''(\qq^2)$ encode the corrections beyond the leading SD coupling to the transverse and longitudinal SD responses, respectively.
They capture the combined effect of the pseudoscalar form factor, radius corrections, and two-body currents. They are given by
\begin{align}
\label{deltas}
\delta'(\qq^2)&= -\frac{\qq^2\,\langle r_A^2\rangle}{6} + \delta a(\qq^2) ,\\
\delta''(\qq^2)&= -\frac{g_{\pi NN}\,F_{\pi}}{g_A\,\mN}
\frac{\qq^2}{\qq^2+M_\pi^2}
+ \delta a(\qq^2) + \delta a^P(\qq^2),\notag
\end{align}
where the two-body current contributions $\delta a(\qq^2)$ and $\delta a^P(\qq^2)$ are defined in Eqs.~\eqref{deltaa}--\eqref{deltaap}.

Note that currents proportional to $(\qq \cdot \sig_i)\qq $ only contribute to the longitudinal multipoles. Moreover, their contribution can be treated similarly to terms proportional to $\sig_i$ because
\begin{align}
\label{qsigmaq_long}
(\qq \cdot \sig_i)\qq = \qq^2 \sig_i + \qq \times (\qq \times \sig_i),
\end{align}
where the second term is perpendicular to $\qq$ and vanishes for longitudinal multipoles.

\begin{figure}[t]
	\begin{center}
	\includegraphics[width=0.48\textwidth,clip]{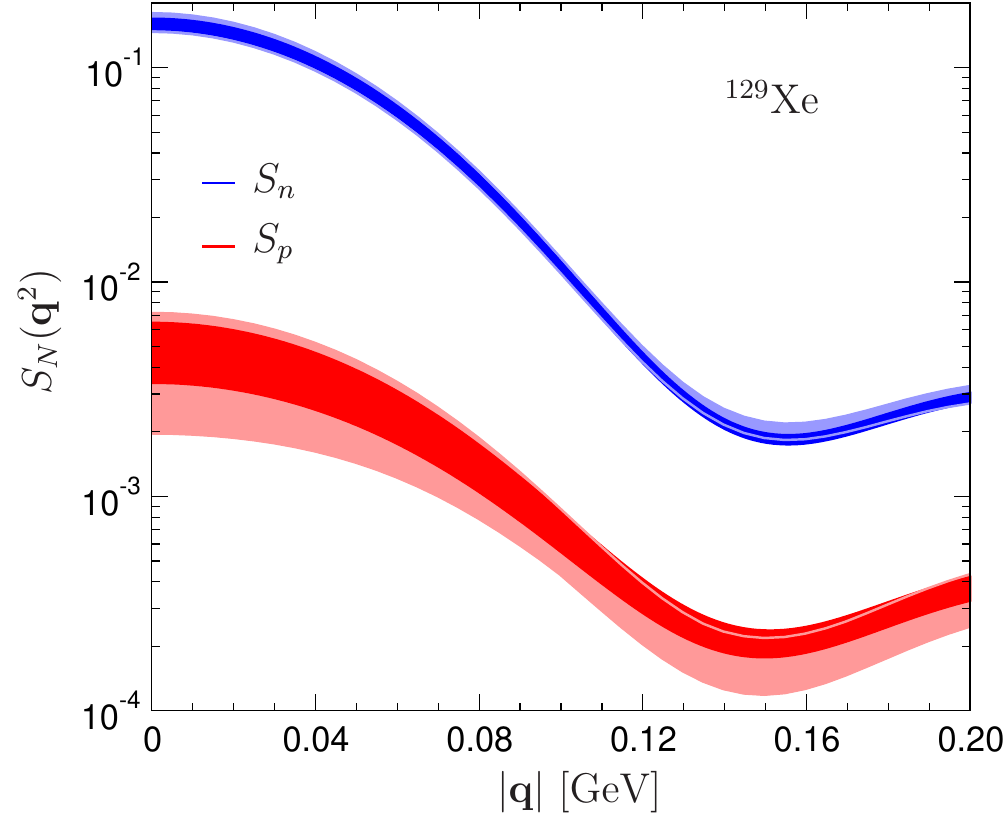}
	\includegraphics[width=0.48\textwidth,clip]{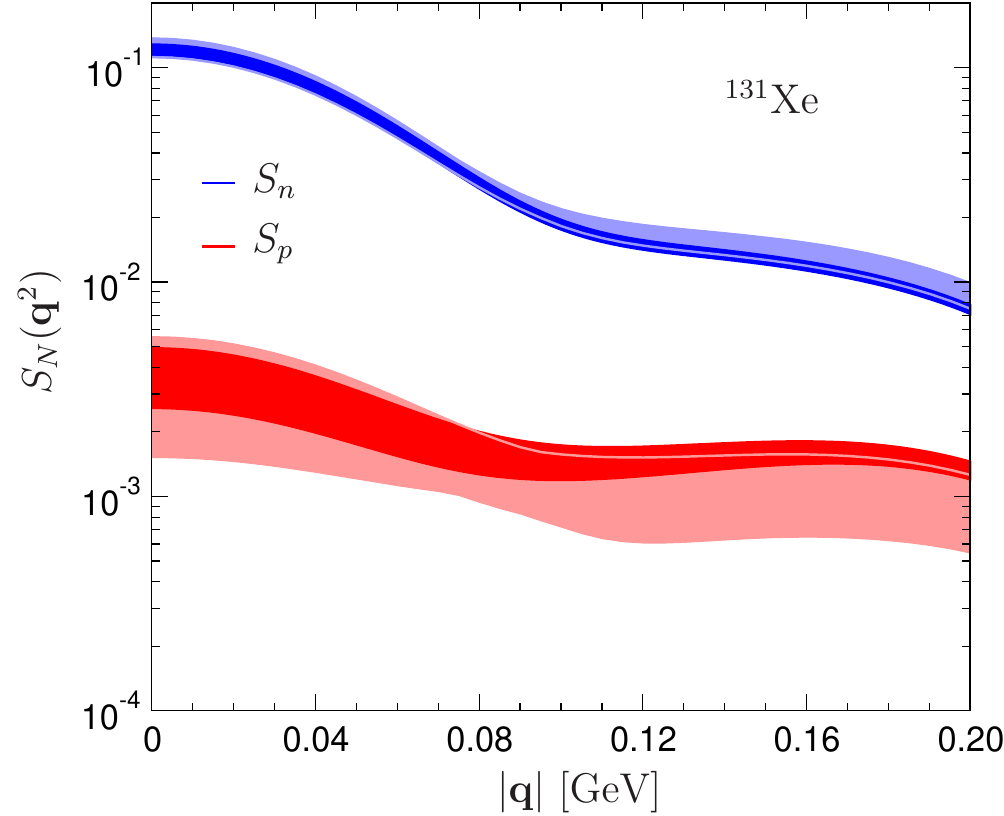}
	\end{center}
\caption{Structure factors $S_N(\qq^2)$, as defined in Eqs.~\eqref{Sp}--\eqref{Sn}, for xenon. The dark bands refer to the results from this work, the light bands to the ones from Ref.~\cite{Klos:2013rwa}.}
\label{fig:Sn_Xe}
\end{figure}

As a first application we show the results for the structure factors $S_N(\qq^2)$ for xenon, in comparison to our previous work from Ref.~\cite{Klos:2013rwa}, see Fig.~\ref{fig:Sn_Xe}. There is good consistency within the earlier theoretical band.
As expected, recent progress in the understanding of low-energy constants and two-body currents in $\beta$ decays allows us to reduce the theoretical uncertainties.
Figure~\ref{fig:Sn_Xe} shows that for xenon this is especially the case for $S_p$, as this response is dominated by two-body contributions.
In general, uncertainty bands are reduced most for the smaller structure factors corresponding to the species with an even number of nucleons.

\begin{figure}[t]
	\begin{center}
	\includegraphics[width=0.48\textwidth,clip]{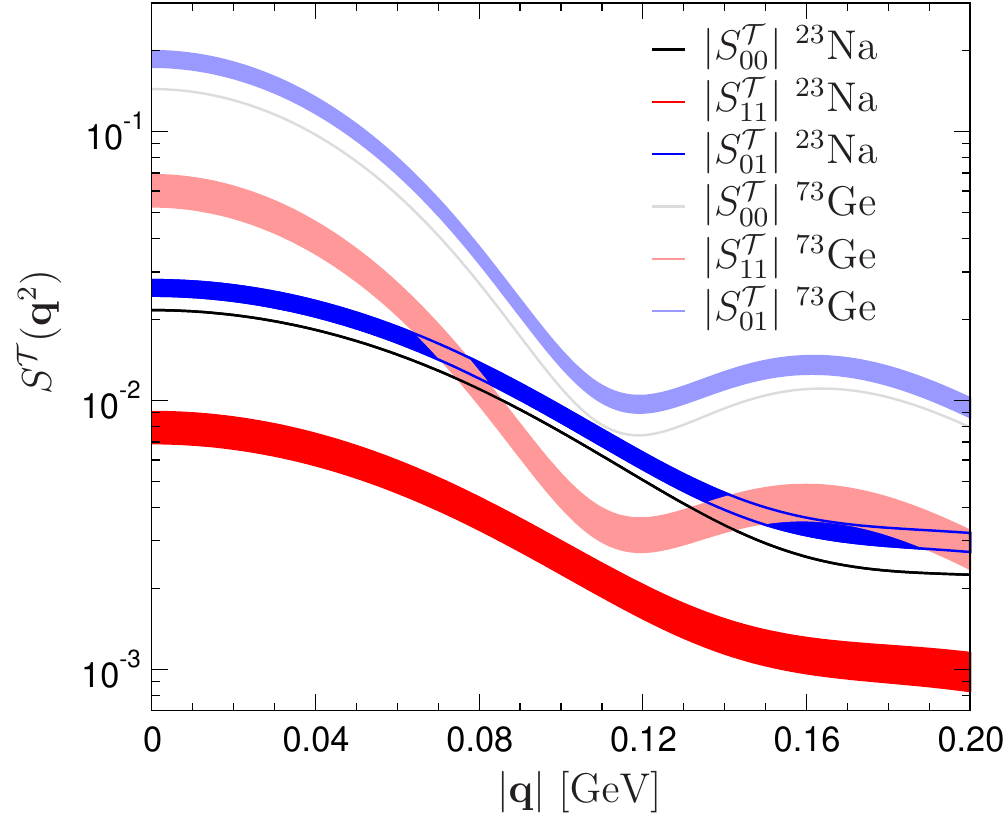}
	\includegraphics[width=0.48\textwidth,clip]{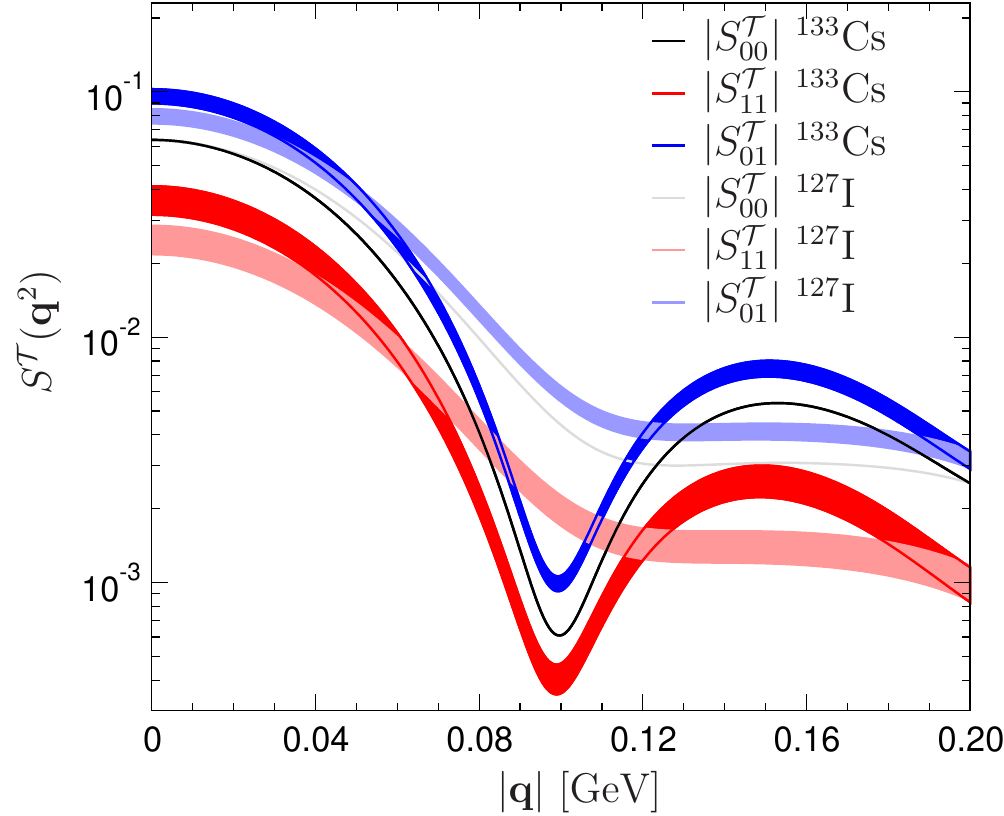}
	\end{center}
\caption{Tranverse SD structure factors for CE$\nu$NS, as required for Eq.~\eqref{CEvNS_SM}. The figure includes all isospin channels, for sodium and germanium (top) and cesium and iodine (bottom).}
\label{fig:SijT}
\end{figure}

\begin{figure}[t]
	\begin{center}
	\includegraphics[width=0.48\textwidth,clip]{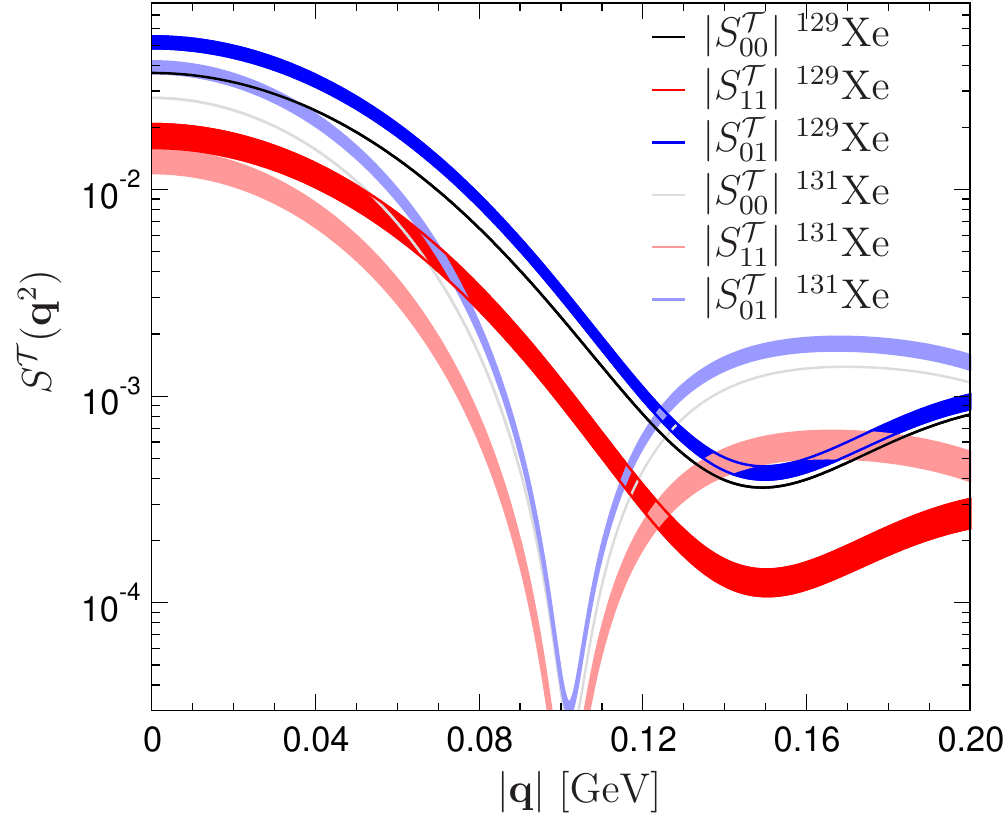}
	\end{center}
\caption{Same as Fig.~\ref{fig:SijT}, for the two odd-mass xenon isotopes.}
\label{fig:SijT_Xe}
\end{figure}

Second, we show the variant of the SD structure factors required for CE$\nu$NS, see Figs.~\ref{fig:SijT} and~\ref{fig:SijT_Xe}. As discussed in Sec.~\ref{sec:CEvNS}, only the transverse multipoles contribute to the final expression in Eq.~\eqref{CEvNS_SM}, but unless the strangeness contribution is neglected all isospin components enter. The figures show our shell-model results, including two-body currents and form factor corrections represented by $\delta'(\qq^2)$, $\delta''(\qq^2)$ in Eq.~\eqref{deltas}. 
For a given nucleus, the shape of the isovector and isoscalar responses is similar because all of them are ultimately dominated by either $S_p(\qq^2)$, if the nucleus has an unpaired proton, or $S_n(\qq^2)$, for nuclei with odd number of neutrons.
A comparison between the $^{131}$Xe structure factors in Figs.~\ref{fig:Sn_Xe} and~\ref{fig:SijT_Xe} shows that the shape of the transverse component may differ significantly from the total structure factor (dominated by the longitudinal component in that case, see Ref.~\cite{Klos:2013rwa}). 
According to Eq.~\eqref{Fsigmanorm}, the normalization of the transverse
contribution differs by $2/3$ from the sum. Moreover, as can be
seen from Figs.~\ref{fig:SijT} and~\ref{fig:SijT_Xe}, the isovector combination
$S_{11}^{\mathcal T}$, which is most relevant for Eq.~\eqref{CEvNS_SM}, is
the smallest of the isospin components. This is partly because of the
reduction caused by axial-vector two-body currents, which are
isovector, as one-body $S_{11}$ and $S_{00}$ structure factors are
of similar size.

\section{Nuclear responses beyond the Standard Model}
\label{sec:BSM}

\subsection{Vector and axial-vector operators}

As a first step, we generalize Eq.~\eqref{CEvNS_SM} to include scenarios in which still only vector and axial-vector operators are present, but whose Wilson coefficients are allowed to deviate from the SM. Especially the case with BSM contributions only to the vector operators is a frequently studied scenario~\cite{Akimov:2017ade,Akimov:2020pdx}. 

To collect the combination of Wilson coefficients and hadronic matrix elements, we define
\begin{align}
\label{couplings}
 g_{V,i}^{N}(t)&=\sum_{q=u,d,s}C_q^V F_{1}^{q,N}(t), \qquad i\in\{1,2\},\notag\\
 g_A^N(t)&=\sum_{q=u,d,s}C_q^A G_A^{q,N}(t),
\end{align}
as well as the short-hand notation
\begin{align}
 g_V^N&\equiv g_{V,1}^{N}(0),\qquad g_{V,2}^N\equiv g_{V,2}^{N}(0),\qquad g_A^N=g_A^N(0),\notag\\
 g_{V,1}^{N}(t)&=g_V^N+\dot g_V^N t+\Order(t^2), 
\end{align}
where 
\begin{align}
\dot g_{V}^p&=g_V^p\bigg(\frac{\langle r_E^2\rangle^p}{6}-\frac{\kappa^p}{4\mN^2}\bigg)+g_V^n\bigg(\frac{\langle r_E^2\rangle^n}{6}-\frac{\kappa^n}{4\mN^2}\bigg)\notag\\
&+g_V^B\bigg(\frac{\langle r_{E,s}^2\rangle^N}{6}-\frac{\kappa^N_s}{4\mN^2}\bigg),\notag\\
g_{V,2}^{p}&=g_V^p\kappa^p+g_V^n\kappa^n+g_V^B\kappa_s^N,
\end{align}
and the neutron equations follow by $g_V^p=2C_u^V+C_d^V\leftrightarrow g_V^n=C_u^V+2C_d^V$. For the strangeness contribution we have introduced the ``baryon-number'' coupling
\beq
g_V^B=\sum_{q=u,d,s}C_q^V.
\eeq
In the SM, where $C_d^V=C_s^V$, this new coupling coincides with $g_V^n$ and was therefore not needed in Eq.~\eqref{Fw_definition}.
Collecting all terms, the generalization of Eq.~\eqref{CEvNS_SM} becomes
\begin{align}
\frac{\diff \sigma_A}{\diff T}&=\frac{\mA}{2\pi}\bigg(1-\frac{\mA T}{2E_\nu^2}-\frac{T}{E_\nu}\bigg)\tilde Q_\text{w}^2\big|\tilde F_\text{w}(\qq^2)\big|^2\notag\\
&+\frac{\mA}{2\pi}\bigg(1+\frac{\mA T}{2E_\nu^2}-\frac{T}{E_\nu}\bigg)\tilde F_A(\qq^2),
% &+\frac{2\mA}{2J+1}\bigg(2+\frac{\mA T}{E_\nu^2}-\frac{2T}{E_\nu}\bigg)\notag\\
%&\hspace{-12pt}\times\Big(\big(g_A^0\big)^2 S_{00}^\mathcal{T}(\qq^2)+g_A^0 g_A^1 S_{01}^\mathcal{T}(\qq^2)+ \big(g_A^1\big)^2 S_{11}^\mathcal{T}(\qq^2)\Big),\notag
\end{align}
where
\begin{align}
 &\tilde F_A(\qq^2)=\frac{8\pi}{2J+1}\\
 &\hspace{23pt}\times\Big(\big(g_A^0\big)^2 S_{00}^\mathcal{T}(\qq^2)+g_A^0 g_A^1 S_{01}^\mathcal{T}(\qq^2)+ \big(g_A^1\big)^2 S_{11}^\mathcal{T}(\qq^2)\Big).\notag
\end{align}
The isoscalar and isovector couplings for the axial-vector part are defined as in Eq.~\eqref{def_axial_vector}, so that $\tilde F_A\to G_F^2/2\, F_A$ in the SM. Similarly, the new ``weak charge,''
\beq
\tilde Q_\text{w}=Zg_V^p+N g_V^n,
\eeq
reduces to $-G_F/\sqrt{2}\, Q_\text{w}$ in the SM, see Eq.~\eqref{Qweak}, and the new ``weak form factor'' becomes
\begin{align}
 \tilde F_\text{w}(\qq^2)&=\frac{1}{\tilde Q_\text{w}}\bigg[\bigg(g_V^p+\dot g_{V}^p t + \frac{g_V^p+2g_{V,2}^p}{8\mN^2} t\bigg) \F^M_p(\qq^2)\notag\\
&+\bigg(g_V^n+\dot g_{V}^n t + \frac{g_V^n+2g_{V,2}^n}{8\mN^2} t \bigg)\F^M_n(\qq^2)\\
&- \frac{g_V^p+2g_{V,2}^p}{4\mN^2} t \F^{\Phi''}_p(\qq^2)
- \frac{g_V^n+2g_{V,2}^n}{4\mN^2} t \F^{\Phi''}_n(\qq^2)\bigg].\notag
\end{align}
Modifications due to BSM physics thus affect the CE$\nu$NS cross section in two ways: the normalization at $\qq^2=0$ changes, visible as the change in the weak charge, but in addition the weak form factor changes as well, which is due to the fact that $Q_\text{w}$ does not actually factorize, but emerges as a sum of different underlying nuclear responses. Only in special cases in which the shifts in the Wilson coefficients are aligned with the SM, i.e., all coefficients are modified by the same relative factor, would $F_\text{w}(\qq^2)$ remain unaltered. 

\begin{figure}[t]
	\begin{center}
	\includegraphics[width=0.48\textwidth,clip]{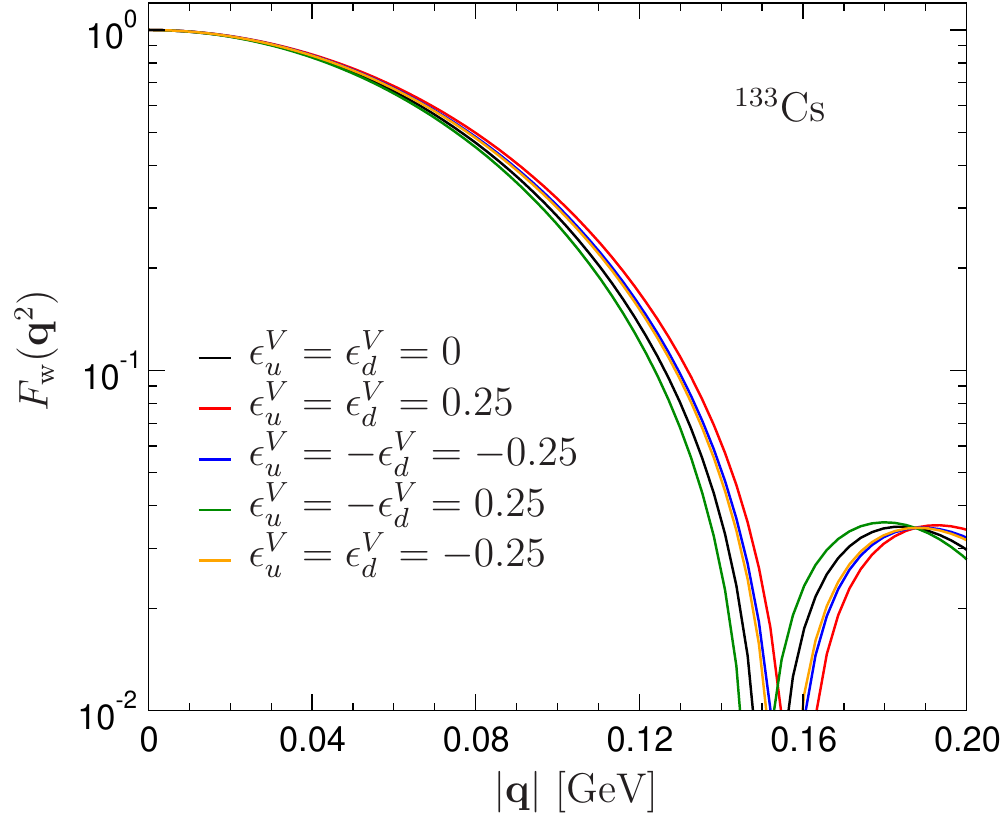}
	\end{center}
\caption{Changes in the weak form factor for $^{133}$Cs in the presence of BSM contributions to the $u$- and $d$-quark vector Wilson coefficients~\eqref{BSM_epsV}.}
\label{fig:Fweak_Cs_BSM}
\end{figure}

\begin{figure}[t]
	\begin{center}
	\includegraphics[width=0.48\textwidth,clip]{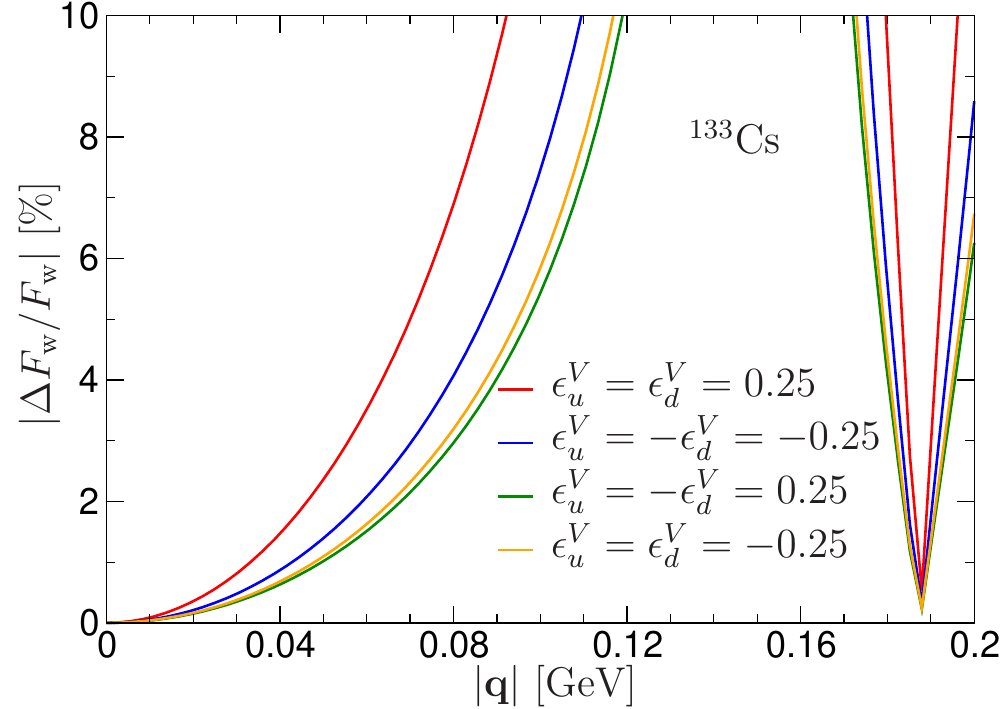}
	\end{center}
\caption{Relative changes in the weak form factor for $^{133}$Cs, for the same scenarios shown in Fig.~\ref{fig:Fweak_Cs_BSM}.}
\label{fig:Fweak_Cs_BSM_rel}
\end{figure}

To quantify the changes with respect to $F_\text{w}(\qq^2)$, the new form factor is shown in Fig.~\ref{fig:Fweak_Cs_BSM} for several points in the BSM parameter space. These contributions to the $u$- and $d$-quark vector Wilson coefficients, defined as in Eq.~\eqref{BSM_epsV}, are large but realistic in view of current bounds from CE$\nu$NS~\cite{Akimov:2017ade,Akimov:2020pdx}. By definition, the deviations vanish at $|\qq|=0$, and they become most visible in the vicinity of the zeros. The second point is illustrated in Fig.~\ref{fig:Fweak_Cs_BSM_rel}, which shows that sufficiently far away from the zeros the changes are at the few-percent level, while the relative deviations are enhanced once the process becomes less coherent.
The relative changes to $F_\text{w}(\qq^2)$ in Fig.~\ref{fig:Fweak_Cs_BSM_rel} are comparable to the current nuclear-structure uncertainties suggested by Fig.~\ref{fig:Fweak_Ar}.  

\subsection{Operators not present in the Standard Model}

Next, we turn to the operators in Eq.~\eqref{operators_basis} not present in the SM. At dimension 5 there is only the dipole operator, leading to the lepton trace
\begin{align}
\label{lepton_trace_dipole}
 L^{\mu\nu}&=-\Tr\big(\slashed{k'}[\gamma^\alpha,\gamma^\mu]P_L\slashed{k}[\gamma^\beta,\gamma^\nu]P_R\big)q_\alpha q_\beta\notag\\
 &=-8t(k^\mu k'^\nu+k'^\mu k^\nu),
\end{align}
where we dropped terms that vanish upon contraction with the nuclear matrix element due to gauge invariance. Since the interference terms with the SM contribution vanish, the presence of a dipole contribution would manifest itself as a new, long-range interaction
\beq
\frac{\diff \sigma_A}{\diff T}\bigg|_\text{dipole}=\frac{4\alpha C_F^2}{T}Z^2\big|F_\text{ch}(\qq^2)\big|^2+\Order(T^0).
\eeq
One power of $1/t$ from the photon propagator in the squared matrix element cancels with the lepton trace in Eq.~\eqref{lepton_trace_dipole}, but the second remains and leads to the divergence for $T\to 0$, due to the relation between momentum transfer and nuclear recoil given in Eq.~\eqref{def_T}. 

Next, the lepton trace for the scalar operator is
\beq
L=\Tr\big(\slashed{k'}P_L\slashed{k}P_R\big)=2k\cdot k'=-t.
\eeq
The diagonal term in the cross section can be expressed as
\beq
\frac{\diff \sigma_A}{\diff T}\bigg|_\text{scalar}=\frac{\mA^2 T}{4\pi E_\nu^2} \big|F_S(\qq^2)\big|^2.
\eeq
This expression vanishes for $T\to 0$, but otherwise there is no kinematic suppression compared to the vector contribution due to the scaling $\mA T/(2E_\nu^2)\lesssim 1$. We have collected all the relevant couplings and form factors in the scalar combination $F_S$, which is defined as
\begin{align}
 F_S(\qq^2)&=\sum_{N=n,p}\Big(f_N+\frac{t}{\mN^2}\dot f_N\Big)\F_N^M(\qq^2)\notag\\
 &+\big(f_\pi+2f_\pi^\theta\Big)\F_\pi(\qq^2)+f_\pi^\theta\F_\text{b}(\qq^2),
\end{align}
with $\F_N^M$ given in Eq.~\eqref{Fpn}, the two-body contributions $\F_\pi(\qq^2)$, $\F_\text{b}(\qq^2)$ from Ref.~\cite{Hoferichter:2018acd}, and the following combinations of Wilson coefficients and hadronic couplings
\begin{align}
 f_N&=\mN\bigg(\sum_{q=u,d,s}C^{S}_{q}f_q^N-12\pi f^N_QC'^S_{g}\bigg),\notag\\
\dot f_N&=C^{S}_{u}\frac{1-\xi_{ud}}{2}\dot\sigma+C^{S}_{d}\frac{1+\xi_{ud}}{2}\dot\sigma+C^{S}_{s}\dot\sigma_s,\notag\\
f_\pi&=\mpi\sum_{q=u,d} \Big(C_{q}^{S}+\frac{8\pi}{9}C'^S_{g}\Big)f_q^\pi,\notag\\
f_\pi^\theta&=-\mpi\frac{8\pi}{9}C'^S_{g}.
\end{align}
Again, there is no interference with the SM, but the scalar contribution does interfere with the dipole, leading to
\begin{align}
\label{dipole_scalar}
\frac{\diff \sigma_A}{\diff T}\bigg|_\text{dipole+scalar}&=\frac{\mA^2 T}{4\pi E_\nu^2} \\
&\times\bigg|F_S(\qq^2)+\frac{2E_\nu-T}{\mA T}Ze C_F F_\text{ch}(\qq^2)\bigg|^2.\notag
\end{align}

For the pseudoscalar operator there is also no interference with the SM, and due to the SD nature of the nucleon matrix elements such a response should be even further suppressed than in the scalar case. To corroborate that expectation we rewrite the operator by means of the axial Ward identity
\beq
\bar\nu P_L\nu\, m_q\bar qi\gamma_5 q = -\frac{i}{2} q_\mu \bar\nu P_L \nu\,\bar q\gamma^\mu\gamma_5 q,
\eeq
so that we can define a leptonic trace
\beq
L^{\mu\nu}=\frac{1}{4}q^\mu q^\nu \Tr\big(\slashed{k'}P_L\slashed{k}P_R\big)=-\frac{t}{4}q^\mu q^\nu,
\eeq
to be contracted with the same nuclear responses already studied for the axial-vector case. The relevant spin sums are given by $L_{33}=L_{ii}=t^2/4$, leading to a kinematic suppression with respect to the axial-vector contribution that scales as 
\beq
\frac{t^2}{16E_\nu^2\mN^2}=\frac{\mA^2T^2}{4E_\nu^2\mN^2}\lesssim \frac{E_\nu^2}{\mN^2}.
\eeq
The scale $\mN$ emerges assuming that the formal difference between the dimension-$7$ and dimension-$6$ operators is mainly due to hadronic scales (as is manifest for the matrix elements of the scalar operator, see Eq.~\eqref{scalar_current}), and for higher scales the suppression would be even stronger. In either case we conclude that pseudoscalar contributions to CE$\nu$NS are negligible. 

For the tensor operator, the most relevant contributions are expected from the space-like components $\sigma_{ij}$, because only those are momentum independent and not suppressed by $1/\mN$ in the nonrelativistic expansion. For the same reason, the induced terms in Eq.~\eqref{tensor_decomposition} are subleading. The result of the multipole decomposition for tensor currents, see App.~\ref{app:multi_pole_decomposition_tensor}, then leads to the following expressions: defining the couplings via
\beq
g^N_{T,1}(t)=\sum_{q=u,d,s}C_q^T F_{1,T}^{q,N}(t),\qquad g^N_{T,1}\equiv g^N_{T,1}(0),
\eeq
and 
\beq
\label{def_tensor_vector}
g_{T,1}^0=\frac{g_{T,1}^p+g_{T,1}^n}{2},\qquad g_{T,1}^1=\frac{g_{T,1}^p-g_{T,1}^n}{2},
\eeq
the cross section becomes
\begin{align}
\frac{\diff \sigma_A}{\diff T}\bigg|_\text{tensor}&=
 \frac{8\mA}{2J+1}\bigg(2-\frac{\mA T}{E_\nu^2}-\frac{2T}{E_\nu}\bigg)\Big[\big(g_{T,1}^0\big)^2 {\bar S}_{00}^\mathcal{T}(\qq^2)\notag\\
 &\qquad+g_{T,1}^0 g_{T,1}^1 {\bar S}_{01}^\mathcal{T}(\qq^2)+ \big(g_{T,1}^1\big)^2 {\bar S}_{11}^\mathcal{T}(\qq^2)\Big]\notag\\
 &+\frac{32\mA}{2J+1}\bigg(1-\frac{T}{E_\nu}\bigg)\Big[\big(g_{T,1}^0\big)^2 {\bar S}_{00}^\mathcal{L}(\qq^2)\\
 &\qquad+g_{T,1}^0 g_{T,1}^1 {\bar S}_{01}^\mathcal{L}(\qq^2)+ \big(g_{T,1}^1\big)^2 {\bar S}_{11}^\mathcal{L}(\qq^2)\Big].\notag
\end{align}
Contrary to the axial-vector response, there is now also a contribution from the longitudinal multipoles, ${\bar S}_{ij}^\mathcal{L}(\qq^2)$. These response functions are identical to the ones derived for the axial-vector case only at leading order, i.e., the two-body corrections for the tensor current would take a different form and likewise the corrections from the induced pseudoscalar
and the axial-vector radius need to be removed
\beq
{\bar S}_{ij}^{\mathcal{T}}(\qq^2) =  S_{ij}^{\mathcal{T}}(\qq^2)\Big|_{\delta'(\qq^2)=0},\quad
{\bar S}_{ij}^{\mathcal{L}}(\qq^2) =  S_{ij}^{\mathcal{L}}(\qq^2)\Big|_{\delta''(\qq^2)=0}.
\eeq

There are again no interference terms with the SM, but the lepton traces do allow for potential interference terms with scalar, pseudoscalar, and dipole operators. In addition, there would be additional contributions from the $\sigma_{0i}$ components of the tensor current as well as the induced form factors in Eq.~\eqref{tensor_decomposition}. In case such contributions became relevant, the formalism could be extended accordingly.

\section{Summary}
\label{sec:summary}

In this paper we have provided a detailed account of the CE$\nu$NS cross section both within the SM and beyond. To this end, we started from a decomposition into effective operators, hadronic matrix elements, and nuclear structure factors, including both the vector and axial-vector operators already present in the SM, but also considering the effects of (pseudo-)scalar, tensor, and dipole operators. Light BSM degrees of freedom could be included along similar lines. 

As a first step, we introduced the charge and weak form factors as typically defined in electron scattering, to exemplify their decomposition in terms of underlying nuclear structure factors, but also hadronic matrix elements and Wilson coefficients. 
The analogous decomposition for CE$\nu$NS is then used to address the question how, e.g., the weak form factor needs to be modified once BSM contributions are permitted, and to derive master formulae for the cross section in the various cases.  

Our results for the nuclear structure factors are based on the large-scale nuclear shell model. In addition to the coherent part of the response, which is largely determined by charge operators, radius and relativistic corrections, as well as spin-orbit contributions, we have also performed a detailed study of the typically neglected axial-vector responses. While the general formalism is similar to the spin-dependent responses for dark matter scattering off nuclei, there are key differences. Most notably, only the transverse multipoles contribute to CE$\nu$NS due to the lepton trace. We have also calculated updates for the structure factors relevant for spin-dependent dark matter scattering.\footnote{Our results for the nuclear structure factors, as can be reconstructed from the fits for the nuclear responses in App.~\ref{app:parameterization}, are also available as text files upon request.}  

Our calculation of the spin-dependent responses takes advantage of several developments in recent years that allow us to improve the treatment of two-body currents as predicted from chiral EFT. These include improved determinations of the relevant low-energy constants from pion--nucleon scattering, the calculation of one-loop corrections to the nuclear axial-vector current, and insights from ab initio studies of two-body effects in medium-mass and heavy nuclei. While the nuclear interactions used in this work are still phenomenological, this strategy allows us to incorporate as many constraints from chiral EFT as possible, including, for the first time, the effect of contact operators and pion-pole contributions to the two-body currents. 

Finally, we provide further details of the multipole expansion of the nuclear responses, tailored towards the aspects relevant for the CE$\nu$NS application and making the connection to the notation in the nuclear-physics literature. Together with the fits of the resulting nuclear responses as well as the EFT decomposition of the cross section, this defines general CE$\nu$NS responses for a wide range of isotopes and effective operators. 

Future precision studies of CE$\nu$NS will require improved nuclear
responses, especially those involving neutrons. As CE$\nu$NS may, in fact,
be the most promising probe of the neutron responses of atomic nuclei,
a global analysis of multiple targets will be required to disentangle
nuclear-structure and potential BSM effects.

\begin{acknowledgments}
We thank C.~Alexandrou, S.~Bacca, A.~Crivellin, J.~Detwiler, J.~Erler, D.~Gazit, H.~Krebs, S.~Pastore, J.~Piekarewicz, K.~Scholberg, A.~Shindler, and J.~de Vries for valuable discussions.  
This work was supported in part by the SNSF (Project No.\ PCEFP2\_181117), the US DOE (Grant No.\ DE-FG02-00ER41132),
the Spanish MICINN through the ``Ram\'on y
Cajal'' program with grant RYC-2017-22781,
the Spanish MINECO grant FIS2017-87534-P,
the Japanese Society for the Promotion of Science through Grant 18K03639,
MEXT as Priority Issue on Post-K Computer (Elucidation of the
Fundamental Laws and Evolution of the Universe), 
%the Joint Institute for Computational Fundamental Science (JICFuS), 
the JICFuS,
the CNS-RIKEN joint project for large-scale nuclear structure calculations, the Deutsche Forschungsgemeinschaft (DFG, German Research Foundation) -- {Project-ID} 279384907 -- SFB 1245, and the Max Planck Society.
\end{acknowledgments}

\newpage

\appendix

\begin{widetext}
 
\section{Multipole expansion}
\label{app:multipole}

In this appendix we review the main features of the multipole expansion, following closely Refs.~\cite{Serot:1978vj,Walecka:1995mi}. The starting point is the leptonic current $l_\mu$, which is decomposed into the temporal component $l_0$ and the spatial, spherical components $l_\lambda$, $\lambda=\pm, 3$ with respect to the reference vector $\qq$, where the latter index is chosen to avoid confusion with the temporal component. The spin sum takes the form   
\begin{align}
\label{multipole_expansion}
 \sum_\text{spins}\big|\langle f|\Lagr|i\rangle\big|^2
 &=4\pi\sum_\text{spins}\bigg(\sum_{L\geq 0}\bigg[l_3 l_3^*\big|\langle J_f||\Lagr_L+\Lagr_L^5||J_i\rangle\big|^2
+l_0l_0^* \big|\langle J_f||\M_L+\M_L^5||J_i\rangle\big|^2\notag\\
&-2\,\Re\Big(l_3l_0^*\langle J_f||\Lagr_L+\Lagr_L^5||J_i\rangle \langle J_f||\M_L+\M_L^5||J_i\rangle^*\Big)\bigg]\notag
 \\
 &+\frac{1}{2}\sum_{\lambda=\pm 1}l_\lambda l_\lambda^*\sum_{L\geq 1}\big|\langle J_f||{\mathcal T}_L^\text{el}+{\mathcal T}_L^\text{el5}+\lambda \big({\mathcal T}_L^\text{mag}+{\mathcal T}_L^\text{mag5}\big)||J_i\rangle\big|^2\bigg),
\end{align}
where the reduced matrix elements refer to the longitudinal ($\Lagr$), Coulomb ($\M$), transverse electric (${\mathcal T}^\text{el}$), and transverse magnetic (${\mathcal T}^\text{mag}$) multipoles. The latter can be simplified to 
 \begin{align}
 \label{multipole_expansion_1}
 \sum_\text{spins}\big|\langle f|\Lagr|i\rangle\big|^2\bigg|_\mathcal{T}&=2\pi\sum_\text{spins}\sum_{L\geq 1}\bigg[(\llvec\cdot \llvec^*-l_3 l_3^*)
 \Big(\big|\langle J_f||{\mathcal T}_L^\text{el}+{\mathcal T}_L^\text{el5}||J_i\rangle\big|^2 + \big|\langle J_f||{\mathcal T}_L^\text{mag}+{\mathcal T}_L^\text{mag5}||J_i\rangle\big|^2\Big)\notag\\
 &-2i(\llvec\times \llvec^*)_3 \Re\Big(\langle J_f||{\mathcal T}_L^\text{el}+{\mathcal T}_L^\text{el5}||J_i\rangle\langle J_f||{\mathcal T}_L^\text{mag}+{\mathcal T}_L^\text{mag5}||J_i\rangle^*\Big)\bigg].
\end{align}
The single-nucleon contributions, obtained by nonrelativistic expansion of Eqs.~\eqref{vector_FF_def} and~\eqref{axial_vector_FF_def}, can then be expressed in terms of fundamental multipole operators according to
\begin{align}
\label{multipoles_matching}
 \M_{LM}&=F_1^N M^M_L+\frac{\qq^2}{4\mN^2}\big(F_1^N+2F_2^N\big)\Big(\Phi''^M_L-\frac{1}{2}M_{LM}\Big),\notag\\
 \Lagr_{LM}&=\frac{q^0}{|\qq|}\,\M_{LM},\notag\\
 {\mathcal T}_{LM}^\text{el}&=\frac{|\qq|}{\mN}\bigg[F_1^N\Delta'^M_L+\frac{F_1^N+F_2^N}{2}\Sigma_L^M\bigg],\notag\\
 {\mathcal T}_{LM}^\text{mag}&=-i\frac{|\qq|}{\mN}\bigg[F_1^N{\Delta}_L^M-\frac{F_1^N+F_2^N}{2}\Sigma'^M_L\bigg],\notag\\
 \M_{LM}^5&=-i\frac{|\qq|}{\mN}G_A^N\Big[\Omega_L^M+\frac{1}{2}\Sigma''^M_L\Big],\notag\\
 \Lagr_{LM}^5&=i\bigg(G_A^N\bigg(1-\frac{\qq^2}{8\mN^2}\bigg)-\frac{\qq^2}{4\mN^2} G_P^N\bigg)\Sigma''^M_L,\notag\\
 {\mathcal T}_{LM}^\text{el5}&=iG_A^N\bigg(1-\frac{\qq^2}{8\mN^2}\bigg)\Sigma'^M_L,\notag\\
 {\mathcal T}_{LM}^\text{mag5}&=G_A^N\bigg(1-\frac{\qq^2}{8\mN^2}\bigg){\Sigma}_L^M,
\end{align}
where we dropped the quark labels for the form factors, terms suppressed by $q^0/\mN$, and several subleading multipoles in the axial-vector contribution. The explicit expressions for the multipoles in harmonic oscillator basis are given in Ref.~\cite{Serot:1978vj}, where an additional operator $\Omega'_L=\Delta''_L-\Phi''_L$ is introduced. Not all multipoles will be needed in the analysis, the most important ones are $M$ and $\Phi''$ for the vector responses and $\Sigma'$, $\Sigma''$ for the SD ones.
The nuclear responses ${\Sigma}_L$, $\Delta'_L$, as well as the combinations $\left(\Delta''_L-\frac{1}{2}M_L\right)$, $\left(\Omega_L+\frac{1}{2}\Sigma''_L\right)$ vanish for elastic scattering.

\section{Nuclear responses}
\label{app:nuc_responses}

The nuclear responses associated to the $M$, $\Sigma'$, $\Sigma''$, and $\Phi''$ operators are defined as
\begin{align}
M_{J}&=\sum_{i} j_J(qr_i)Y_{J}(\hat{\rr}_i), \\
\Sigma'_{J}&=  \sum_{i} \frac{1}{\sqrt{2J+1}}\Bigl[-\sqrt{J}\,
j_{J+1}(qr_i) \bigl[Y_{J+1}(\hat{\rr}_i){ \sig}_i\bigr]^J
+\sqrt{J+1}\, 
j_{J-1}(qr_i) \bigl[Y_{J-1}(\hat{\rr}_i){ \sig}_i\bigr]^J
\Bigr], \label{Sigma'} \\
\Sigma''_{J}&= \sum_{i} \frac{1}{\sqrt{2J+1}}
\Bigl[\sqrt{J+1}\,
j_{J+1}(qr_i) \bigl[Y_{J+1}(\hat{\rr}_i){ \sig}_i\bigr]^J
+\sqrt{J}\,
j_{J-1}(qr_i) \bigl[Y_{J-1}(\hat{\rr}_i){ \sig}_i\bigr]^J
\Bigr] \,, \label{Sigma''} \\
\Phi''_{J}&=i\sum_{i}\frac{1}{q}\bfnabla_i \big(j_J(qr_i)Y_{J}(\hat{{\bf r_i}})\big)
\cdot\left({\sig}_i\times\frac{1}{q}\bfnabla_i\right) \nonumber \\
&=i\sum_{i}
\frac{\sqrt{J+1}}{\sqrt{2J+1}}
\left[j_{J+1}(qr)Y_{J+1}(\hat{\rr}_i)
\left({\sig}_i\times\frac{1}{q}\bfnabla_i\right)\right]^J
+
\frac{\sqrt{J}}{\sqrt{2J+1}}
\left[j_{J-1}(qr)Y_{J-1}(\hat{\rr}_i)
\left({\sig}_i\times\frac{1}{q}\bfnabla_i\right)\right]^J,
\end{align}
where $[O_1\,O_2]^J$ indicates the coupling of operators $O_1$ and $O_2$ to a tensor of rank $J$, and tensor projections are omitted.
The single-particle harmonic-oscillator matrix elements needed for the calculation of the nuclear responses are
\begin{align}
\Big\langle n'l'\frac{1}{2}j'\Big|\Big|M_{J}\Big|\Big|nl\frac{1}{2}j\Big\rangle
&=\bra{n'l'}j_{J}(qr_i)\ket{nl} (-1)^{j+1/2+J}
\sqrt{\frac{1}{4\pi}} \, \bigl[(2j'+1)(2j+1)\bigr]^{\frac{1}{2}}
\bigl[(2J+1)(2l+1)(2l'+1)\bigr]^\frac{1}{2} \nonumber \\
&\quad\times \left(
\begin{array}{ccc}
l' & J & l \\
0 & 0 & 0
\end{array}\right)			
\left\lbrace
\begin{array}{ccc}
l' & j' & \nicefrac{1}{2} \\
j & l & J
\end{array}
\right\rbrace,
\end{align}
\begin{align}
\Big\langle n'l'\frac{1}{2}j'\Big|\Big|j_{J'}(pr_i) \bigl[Y_{J'}(\hat{\rr}_i){ \sig}_i\bigr]^J\Big|\Big|nl\frac{1}{2}j\Big\rangle
&=\bra{n'l'}j_{J'}(qr_i)\ket{nl} (-1)^{l'}
\sqrt{\frac{6}{4\pi}} \, \bigl[(2l'+1)(2l+1)(2j'+1)(2j+1)\bigr]^{\frac{1}{2}}\notag\\
&\times \bigl[(2J'+1)(2J+1)\bigr]^\frac{1}{2} 
\left(
\begin{array}{ccc}
l' & J' & l \\
0 & 0 & 0
\end{array}\right)			
\left\lbrace
\begin{array}{ccc}
l' & l & J' \\
\nicefrac{1}{2} & \nicefrac{1}{2} & 1 \\
j' & j & J
\end{array}\right\rbrace,
\end{align}
\begin{align}
\langle n' l' j'||\Phi''_J||n l j\rangle&=
(-1)^{l'}\frac{6}{\sqrt{4\pi}}\sqrt{(2j'+1)(2j+1)(2l'+1)}  \\
&\times\left\lbrace
\sqrt{(J+1)(2J+3)}
\sum_{L=J}^{J+1}(-1)^{J+L}(2L+1)
\Bigg\lbrace
\begin{matrix}                                                
J+1 & 1 & L \\
1 & J & 1
\end{matrix}
\right\rbrace
\left\lbrace
\begin{matrix}                                                
l' & l & L \\
\nicefrac{1}{2} & \nicefrac{1}{2} & 1 \\
j'  & j   & J
\end{matrix}
\right\rbrace \nonumber \\
&\quad\times\bigg[
\sqrt{(l+1)(2l+3)} 
\left\lbrace
\begin{matrix}                                                
J+1 & 1 & L \\
l & l' & l+1
\end{matrix}
\right\rbrace
\left(
\begin{matrix}                                                
l' & J+ 1 & l+1 \\
0 & 0 & 0
\end{matrix}
\right)
\Big\langle n' l'\Big|j_{J+1}(qr_i)
\left(\frac{\partial}{\partial(qr_i)}-\frac{l}{qr_i}\right) \Big|n l\Big\rangle 
\nonumber \\
&\quad\quad-
\sqrt{l(2l-1)} 
\left\lbrace
\begin{matrix}                                                
J+1 & 1 & L \\
l & l' & l-1
\end{matrix}
\right\rbrace
\left(
\begin{matrix}                                                
l' & J+ 1 & l-1 \\
0 & 0 & 0
\end{matrix}
\right)
\Big\langle n' l'\Big|j_{J+1}(qr_i)
\left(\frac{\partial}{\partial(qr_i)}+\frac{l+1}{qr_i}\right) \Big|n l\Big\rangle 
\bigg]  \nonumber \\
&+
\sqrt{J(2J-1)}
\sum_{L=J-1}^{J}(-1)^{J+L}(2L+1)
\left\lbrace
\begin{matrix}                                                
J-1 & 1 & L \\
1 & J & 1
\end{matrix}
\right\rbrace
\left\lbrace
\begin{matrix}                                                
l' & l & L \\
\nicefrac{1}{2} & \nicefrac{1}{2} & 1 \\
j'  & j   & J
\end{matrix}
\right\rbrace \nonumber \\
&\quad\times\bigg[
\sqrt{(l+1)(2l+3)} 
\left\lbrace
\begin{matrix}                                                
J-1 & 1 & L \\
l & l' & l+1
\end{matrix}
\right\rbrace
\left(
\begin{matrix}                                                
l' & J-1 & l+1 \\
0 & 0 & 0
\end{matrix}
\right)
\Big\langle n' l'\Big|j_{J-1}(qr_i)
\left(\frac{\partial}{\partial(qr_i)}-\frac{l}{qr_i}\right) \Big|n l\Big\rangle 
\nonumber \\
&\qquad\quad- 
\sqrt{l(2l-1)} 
\left\lbrace
\begin{matrix}                                                
J-1 & 1 & L \\
l & l' & l-1
\end{matrix}
\right\rbrace
\left(
\begin{matrix}                                                
l' & J-1 & l-1 \\
0 & 0 & 0
\end{matrix}
\right)
\Big\langle n' l'\Big|j_{J-1}(qr_i)
\left(\frac{\partial}{\partial(qr_i)}+\frac{l+1}{qr_i}\right) \Big|n l\Big\rangle 
\bigg]
\Bigg\rbrace.\notag
\end{align}

\section{Nuclear structure calculation of $^{133}$Cs}
\label{app:133cs}
		
In order to illustrate the quality of the shell-model calculations for $^{133}$Cs, Fig.~\ref{fig:cs133_spectrum} compares the calculated and experimental low-energy excitation spectrum of $^{133}$Cs.
Even though our calculation incorrectly predicts a ground state with angular momentum and parity $J^P=5/2^+$, the difference with the $7/2^+$ state is only 10~keV.
The angular momentum and parity of the lowest energy levels is predicted well, even though the energy of the calculated second $3/2^+$ state is lower than in experiment. 
Overall the agreement with experiment is similar as in other odd-mass nuclei with similar mass number.
		
\begin{figure}[t]
	\begin{center}
	\includegraphics[width=0.48\textwidth,clip]{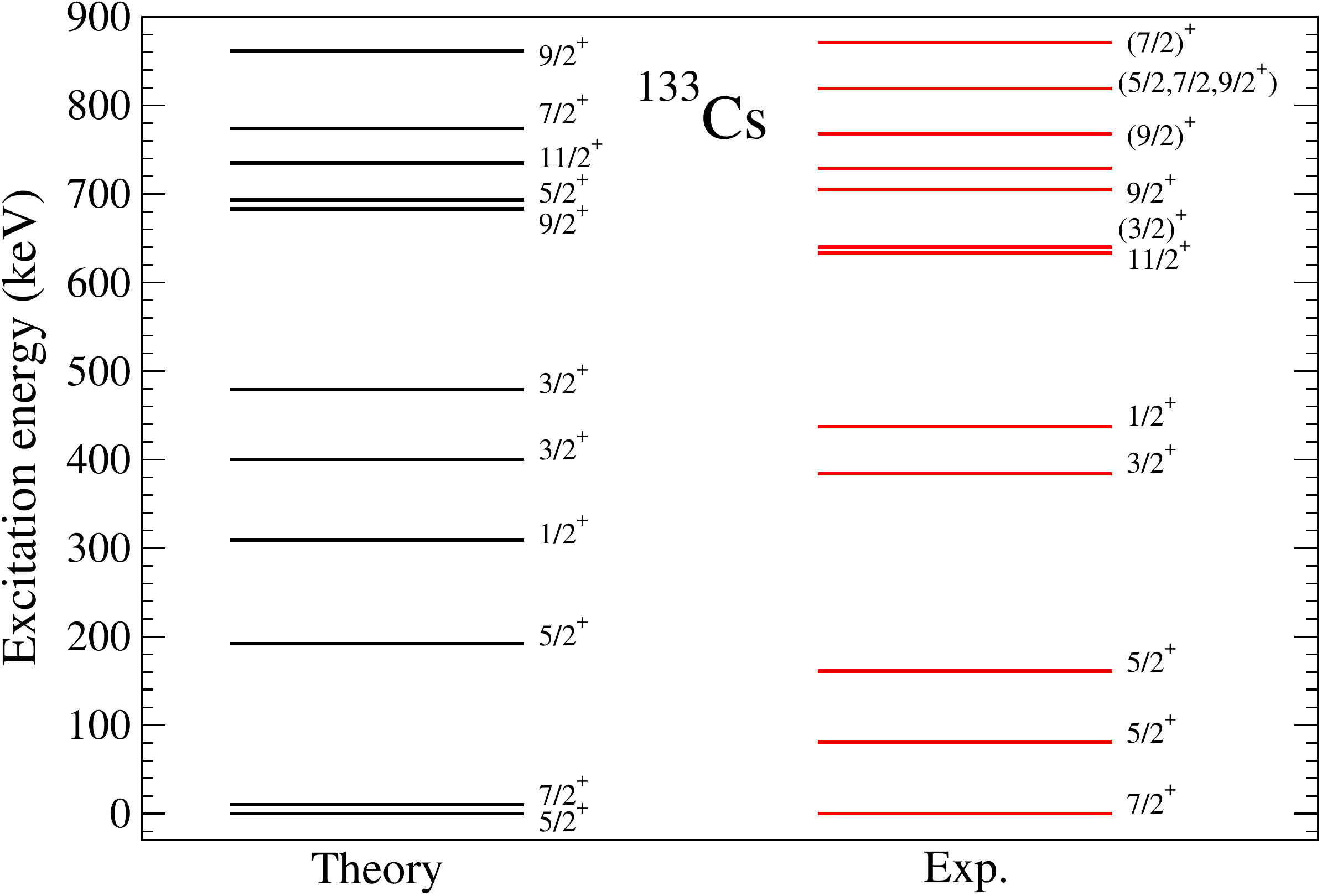}
	\end{center}
\caption{Calculated $^{133}$Cs spectrum compared to experiment.\label{fig:cs133_spectrum}}
\end{figure}

 \section{Multipole decomposition for tensor currents}
 \label{app:multi_pole_decomposition_tensor}
 
Including the tensor operator from Eq.~\eqref{operators_basis} into the analysis requires a generalization of the multipole decomposition reviewed in App.~\ref{app:multipole}. 
Here we follow closely the original derivation in
Refs.~\cite{Glick-Magid:2016,Glick-Magid:2016rsv,Glick-Magid:2020}, including the lepton trace
\begin{align}
 L^{\mu\nu\lambda\sigma}&=\Tr\big(\slashed{k'}\sigma^{\mu\nu} P_L\slashed{k}\sigma^{\lambda\sigma} P_R\big)\notag\\
 &=2\bigg[\big(g^{\mu\lambda}g^{\nu\sigma}-g^{\mu\sigma}g^{\nu\lambda}\big)k\cdot k'
 +i\eps^{\mu\nu\lambda\alpha}k^\sigma k'_\alpha
 -i\eps^{\mu\nu\sigma\alpha}k^\lambda k'_\alpha
 - i \eps^{\mu\lambda\sigma\alpha}k'^\nu k_\alpha
 + i \eps^{\nu\lambda\sigma\alpha}k'^\mu k_\alpha\notag\\
 &-g^{\mu\lambda}\big(k^\nu k'^\sigma+k'^\nu k^\sigma\big)+g^{\mu\sigma}\big(k^\nu k'^\lambda+k'^\nu k^\lambda\big)+g^{\nu\lambda}\big(k^\mu k'^\sigma+k'^\mu k^\sigma\big)-g^{\nu\sigma}\big(k^\mu k'^\lambda+k'^\mu k^\lambda\big)
 \bigg],
\end{align}
and then specify the spin sums relevant for CE$\nu$NS.
The key idea in the generalized multipole expansion is then that the antisymmetric tensor current $j^{\mu\nu}$ essentially admits two vectorial components, $j_i^{(0)}=j_{0i}$ and $j_i^{(1)}=-\frac{i}{\sqrt{2}}\eps_{ijk}j_{jk}$, in terms of which the analog of Eqs.~\eqref{multipole_expansion} and~\eqref{multipole_expansion_1}  becomes~\cite{Glick-Magid:2016,Glick-Magid:2020} 
\begin{align}
\label{multipole_expansion_tensor}
 \sum_\text{spins}\big|\langle f|\Lagr|i\rangle\big|^2
 &=4\pi\sum_\text{spins}\sum_{L\geq 0}\bigg[l_3^{(1)} l_3^{(1)*}\big|\langle J_f||\Lagr_L^{(1)}||J_i\rangle\big|^2+4l_3^{(0)} l_3^{(0)*} \big|\langle J_f||\Lagr_L^{(0)}||J_i\rangle\big|^2\notag\\
 &
+4\,\Re\Big(l_3^{(1)}l_3^{(0)*}\langle J_f||\Lagr_L^{(1)}||J_i\rangle \langle J_f||\Lagr_L^{(0)}||J_i\rangle^*\Big)\bigg]\notag
 \\
 &+2\pi\sum_\text{spins}\sum_{L\geq 1}\bigg[\big(\llvec^{(1)}\cdot \llvec^{(1)*}-l_3^{(1)} l_3^{(1)*}\big)
 \Big(\big|\langle J_f||{\mathcal T}_L^\text{el\,(1)}||J_i\rangle\big|^2 + \big|\langle J_f||{\mathcal T}_L^\text{mag\,(1)}||J_i\rangle\big|^2\Big)\notag\\
 &+4\big(\llvec^{(0)}\cdot \llvec^{(0)*}-l_3^{(0)} l_3^{(0)*}\big)
 \Big(\big|\langle J_f||{\mathcal T}_L^\text{el\,(0)}||J_i\rangle\big|^2 + \big|\langle J_f||{\mathcal T}_L^\text{mag\,(0)}||J_i\rangle\big|^2\Big)\notag\\
 &+4\big(\llvec^{(1)}\cdot \llvec^{(0)*}-l_3^{(1)} l_3^{(0)*}\big)
 \Big(\langle J_f||{\mathcal T}_L^\text{el\,(1)}||J_i\rangle \langle J_f||{\mathcal T}_L^\text{el\,(0)}||J_i\rangle^* + \langle J_f||{\mathcal T}_L^\text{mag\,(1)}||J_i\rangle \langle J_f||{\mathcal T}_L^\text{mag\,(0)}||J_i\rangle^*\Big)\notag\\
 &-2i\big(\llvec^{(1)}\times \llvec^{(1)*}\big)_3 \Re\Big(\langle J_f||{\mathcal T}_L^\text{el\,(1)}||J_i\rangle\langle J_f||{\mathcal T}_L^\text{mag\,(1)}||J_i\rangle^*\Big)\notag\\
 &-8i\big(\llvec^{(0)}\times \llvec^{(0)*}\big)_3 \Re\Big(\langle J_f||{\mathcal T}_L^\text{el\,(0)}||J_i\rangle\langle J_f||{\mathcal T}_L^\text{mag\,(0)}||J_i\rangle^*\Big)\notag\\
 &-4i\big(\llvec^{(1)}\times \llvec^{(0)*}\big)_3 \Re\Big(\langle J_f||{\mathcal T}_L^\text{el\,(1)}||J_i\rangle\langle J_f||{\mathcal T}_L^\text{mag\,(0)}||J_i\rangle^*+\langle J_f||{\mathcal T}_L^\text{mag\,(1)}||J_i\rangle\langle J_f||{\mathcal T}_L^\text{el\,(0)}||J_i\rangle^*\Big)
 \bigg],
\end{align}
where we dropped the distinction between the two parities in each multipole. Since the nonrelativistic reduction of $\sigma_{0i}$ only starts at $\Order(1/\mN)$ and depends on momenta, the most interesting tensor contribution originates from the $\sigma_{ij}\to\eps_{ijk}\sigma_k$ components, contained in $\mathbf{j}^{(1)}$. The relevant spin sum reads
\beq
\sum_\text{spins}l_i^{(1)} l_j^{(1)*}=\frac{1}{2}\eps_{ikl}\eps_{jmn} L_{klmn}
=-2t\delta_{ij}+4\big(\delta_{ij}\kk\cdot \kk'-k_ik'_j-k'_ik_j\big)
+4i\eps_{ijk}\big(k_kE_\nu' +k_k'E_\nu\big),
\eeq
with projections
\begin{align}
 \sum_\text{spins}l_3^{(1)} l_3^{(1)*}&=8E_\nu^2\bigg(1-\frac{T}{E_\nu}\bigg),\notag\\
  \sum_\text{spins}\big(\llvec^{(1)}\cdot \llvec^{(1)*}-l_3^{(1)} l_3^{(1)*}\big)&=4E_\nu^2\bigg(2-\frac{\mA T}{E_\nu^2}-\frac{2T}{E_\nu}\bigg),\notag\\
  \sum_\text{spins}\big(\llvec^{(1)}\times \llvec^{(1)*}\big)_3&=-8iE_\nu^2\sqrt{\frac{2T}{\mA}}.
\end{align}  
In contrast to Eq.~\eqref{spin_sums}, the longitudinal multipole is no longer kinematically suppressed, but instead the interference term between electric and magnetic multipoles can be dropped. In our normalization the hadronic current starts with $-\frac{i}{\sqrt{2}}\eps_{ijk}\sigma_{jk}\to -i\sqrt{2}\sigma_i$, so that, up to the prefactor and the different lepton traces, the remainder of the calculation follows along the same lines as for the axial-vector response.  
 
\section{Parameterizations of the nuclear responses}
\label{app:parameterization}

In this appendix we provide explicit parameterizations for the $M$ and $\Phi''$ responses not already given in previous work~\cite{Hoferichter:2016nvd}, see Tables~\ref{tab:fits} and \ref{tab:fits_Ge}. The parameterizations for the $\Sigma'$ and $\Sigma''$ responses are given in Tables~\ref{tab:fits_SD_1}--\ref{tab:fits_SD_4}.

\begin{table*}[t]
\centering
\renewcommand{\arraystretch}{1.3}
\begin{tabular}{lccccc}
\toprule
Isotope	& $^{19}$F & $^{23}$Na & $^{40}$Ar & $^{127}$I  & $^{133}$Cs\\
\colrule
$J^P$ & $1/2^+$ & $3/2^+$ & $0^+$ & $5/2^+$ & $7/2^+$\\
\colrule
$b$~[fm] & $1.7623$ & $1.8048$ & $1.9399$ & $2.2821$ & $2.2976$ \\
\colrule
$c_1^{M+}$ & $-6.00039$ & $-8.66651$ & $-20.9778$ & $-125.164$ & $-134.2$\\
$c_2^{M+}$ & $0.317846$ & $0.555305$ & $2.41486$ & $35.3993$ & $38.9577$\\
$c_3^{M+}$ & -- & -- & $-0.0368597$ & $-3.62687$ & $-4.12938$\\
$c_4^{M+}$ & -- & -- & -- & $0.125083$ & $0.151119$\\
$c_5^{M+}$ & -- & -- & -- & $-0.000670162$ & $-0.00103353$\\\colrule
$c_1^{M-}$ & $0.666687$ & $0.666658$ & $3.42422$ & $30.4307$ & $33.9495$\\
$c_2^{M-}$ & $-0.102251$ & $-0.0655647$ & $-0.618209$ & $-12.321$ & $-13.9502$\\
$c_3^{M-}$ & -- & -- & $0.0268957$ & $1.78131$ & $2.04567$\\
$c_4^{M-}$ & -- & -- & -- & $-0.0870947$ & $-0.102733$\\
$c_5^{M-}$ & -- & -- & -- & $0.000697815$ & $0.000944352$\\\colrule
$c_0^{\Phi''+}$ & $-0.764186$ & $-2.89325$ & $-4.79093$ & $-26.1218$ & $-28.2527$\\
$c_1^{\Phi''+}$ & $0.152842$ & $0.578667$ & $1.4068$ & $18.1692$ & $20.4868$\\
$c_2^{\Phi''+}$ & -- & -- & $-0.0683192$ & $-3.50413$ & $-4.09303$\\
$c_3^{\Phi''+}$ & -- & -- & -- & $0.223523$ & $0.275572$\\
$c_4^{\Phi''+}$ & -- & -- & -- & $-0.00360552$ & $-0.0051254$\\\colrule
$c_0^{\Phi''-}$ & $0.36285$ & $0.336942$ & $0.326509$ & $3.58476$ & $8.98993$\\
$c_1^{\Phi''-}$ & $-0.0725723$ & $-0.0673903$ & $-0.452519$ & $-4.58091$ & $-8.67714$\\
$c_2^{\Phi''-}$ & -- & -- & $0.0589909$ & $1.46191$ & $2.21868$\\
$c_3^{\Phi''-}$ & -- & -- & -- & $-0.139708$ & $-0.189453$\\
$c_4^{\Phi''-}$ & -- & -- & -- & $0.0035109$ & $0.00473947$\\\botrule
\end{tabular}
\renewcommand{\arraystretch}{1.0}
\caption{Spin/parity $J^P$ of the nuclear ground states, 
harmonic-oscillator length $b$, and fit coefficients for the nuclear
response functions $\F_\pm^M$ and $\F_\pm^{\Phi''}$.  The fit
functions are $\F_\pm^{M}(u) = e^{-\frac{u}{2}}\sum_{i=0}^{n_M} c_i^{M\pm} u^i$
(with $c_0=A$ and $c_0=Z-N$, respectively) and $\F_{\pm}^{\Phi''}(u) =
e^{-\frac{u}{2}} \sum_{i=0}^{n_{\Phi''}} c_i^{\Phi''\pm} u^i$, with $u=\qq^2b^2/2$.  These
forms correspond to the analytical solution in the harmonic-oscillator
basis~\cite{Donnelly:1979ezn,Haxton:2007qx}, with $n_M$ and $n_{\Phi''}$ as implied by the table. Our results for xenon are given in Ref.~\cite{Hoferichter:2016nvd}, the ones for germanium in Table~\ref{tab:fits_Ge}.}
\label{tab:fits}
\end{table*}

\begin{table*}[t]
\centering
\renewcommand{\arraystretch}{1.3}
\begin{tabular}{lccccc}
\toprule
Isotope	& $^{70}$Ge & $^{72}$Ge & $^{73}$Ge & $^{74}$Ge  & $^{76}$Ge\\
\colrule
$J^P$ & $0^+$ & $0^+$ & $9/2^+$ & $0^+$ & $0^+$\\
\colrule
$b$~[fm] & $2.0952$ & $2.1035$ & $2.1076$ & $2.1117$ & $2.1120$ \\
\colrule
$c_1^{M+}$ & $-51.2373$	& $-53.5901	$ & $-54.7404$ & $	-55.9913	$ & $-58.3541$\\
$c_2^{M+}$  & $ 9.61013	$ & $10.2948	$ & $10.6249	$ & $10.9743	$ & $11.6381$\\
$c_3^{M+}$  & $ -0.515768	$ & $-0.57547	$ & $-0.603598	$ & $-0.634449	$ & $-0.691196$\\
$c_4^{M+}$  & $ 0.0039318	$ & $0.0050503	$ & $0.00552928	$ & $0.00632403	$ & $0.00747821$\\\colrule
$c_1^{M-} $ & $ 6.06953	$ & $ 8.67126 $ & $	9.80348	$ & $ 11.356	$ & $ 13.9175$ \\
$c_2^{M-} $ & $ -1.71276 $ & $	-2.51496 $ & $	-2.84183 $ & $	-3.34586 $ & $	-4.13067$ \\
$c_3^{M-} $ & $ 0.130409 $ & $	0.20692 $ & $	0.234571 $ & $	0.287529 $ & $	0.361556$ \\
$c_4^{M-} $ & $ -2.22453\times 10^{-4} $ & $	-0.00213335 $ & $	-0.00255345	$ & $ -0.0043077 $ & $	-0.00609108$\\\colrule
$c_0^{\Phi''+} $ & $ -14.7388	$ & $-15.3806$ & $	-15.5467	$ & $-16.2171$ & $	-16.7737$\\
$c_1^{\Phi''+} $ & $ 7.10953$ & $	7.53352	$ & $7.6085$ & $	8.07754$ & $	8.59006$\\
$c_2^{\Phi''+} $ & $ -0.811295 $ & $	-0.869702	$ & $-0.875102	$ & $-0.951994	$ & $-1.04772$\\
$c_3^{\Phi''+} $ & $ 0.0193996$ & $ 	0.0219601	$ & $0.0220616	$ & $0.0252548	$ & $0.02986$\\\colrule
$c_0^{\Phi''-} $ & $-3.27309	$ & $-0.924438	$ & $-0.848625	$ & $ 2.04591	$ & $4.22205$\\
$c_1^{\Phi''-} $ & $1.25408	$ & $-0.166778	$ & $-0.302814	$ & $-1.90271	$ & $-3.22233$\\
$c_2^{\Phi''-} $ & $-0.0487671	$ & $0.146851	$ & $0.177212	$ & $0.388221	$ & $0.576533$\\
$c_3^{\Phi''-} $ & $-8.74439\times 10^{-4}	$ & $-0.00851802	$ & $-0.0101962	$ & $-0.017214	$ & $-0.0243454$\\\botrule
\end{tabular}
\renewcommand{\arraystretch}{1.0}
\caption{Same as Table~\ref{tab:fits}, for germanium isotopes.}
\label{tab:fits_Ge}
\end{table*}

\begin{table*}[t]
\centering
\renewcommand{\arraystretch}{1.3}
\begin{tabular}{lcccccc}
\toprule
Isotope	& $^{19}$F & \multicolumn{2}{c}{$^{23}$Na} & $^{129}$Xe & \multicolumn{2}{c}{$^{131}$Xe}\\
$L$ & $1$ & $1$ &$3$ & $1$ & $1$ &$3$\\
\colrule
$c_0^{\Sigma'p}$ & $0.269513$ & $0.132973$ & -- & $0.00576416$ & $-0.00511011$ & --\\
$c_1^{\Sigma'p}$ & $-0.18098$ & $-0.104393$ & $0.0899535$  & $-0.0069211$ & $0.00702863$ & $-0.0000968882$\\
$c_2^{\Sigma'p}$ & $0.0296873$ & $0.00909271$ & $-0.0142746$  & $0.00450247$ & $-0.00156217$ & $0.000171958$\\
$c_3^{\Sigma'p}$ & -- & -- & -- & $-0.000867868$ & $0.0000331178$ & $-0.0000934431$\\
$c_4^{\Sigma'p}$ & -- & -- & -- & $0.000038544$ & $3.08471\times10^{-6}$ & $7.87133\times10^{-6}$\\
$c_5^{\Sigma'p}$ & -- & -- & -- & $9.80727\times 10^{-9}$ & $-1.94585\times10^{-8}$ & $-1.56561\times10^{-8}$\\\colrule
$c_0^{\Sigma'n}$ & $-0.00113172$ & $0.0141201$ & -- & $0.185828$ & $-0.161697$ & --\\
$c_1^{\Sigma'n}$ & $0.00038188$ & $-0.00774151$ & $-0.000878018$  & $-0.267263$ & $0.334948$ & $0.0364067$\\
$c_2^{\Sigma'n}$ & $0.000744991$ & $0.000326936$ & $-0.000231297$  & $0.149565$ & $-0.174187$ & $-0.079646$\\
$c_3^{\Sigma'n}$ & -- & -- & -- & $-0.0274886$ & $0.0310707$& $0.022489$\\
$c_4^{\Sigma'n}$ & -- & -- & -- & $0.00173304$ & $-0.00151254$& $-0.00171746$\\
$c_5^{\Sigma'n}$ & -- & -- & -- & $-3.87392\times10^{-7}$& $-3.84408\times10^{-7}$& $-4.0527\times10^{-7}$\\\colrule
$c_0^{\Sigma''p}$ & $0.190574$ & $0.0940265$ & -- & $0.00407586$ & $-0.00361339$ & --\\
$c_1^{\Sigma''p}$ & $-0.125204$ & $-0.0404172$ & $0.0779019$  & $-0.00646161$ & $0.00442108$ & $-0.0000839117$\\
$c_2^{\Sigma''p}$ & $0.0206132$& $-0.000254736$ & $-0.00592251$  & $0.00321675$ & $-0.00205213$& $0.000213614$\\
$c_3^{\Sigma''p}$ & -- & -- & -- & $-0.000582408$ &$0.000349931$ & $-0.0000258884$\\
$c_4^{\Sigma''p}$ & -- & -- & -- & $0.0000294951$&$-0.0000169039$ & $2.73765\times10^{-7}$\\
$c_5^{\Sigma''p}$ & -- & -- & -- & $3.82107\times10^{-9}$& $-3.20028\times10^{-9}$& $-4.49323\times10^{-9}$\\\colrule
$c_0^{\Sigma''n}$ & $-0.000800244$ & $0.00998438$ & -- & $0.131401$ & $-0.114337$ & --\\
$c_1^{\Sigma''n}$ & $0.00106046$ & $-0.00902057$ & $-0.000760388$  & $-0.150054$ & $-0.0175951$ & $0.0315279$\\
$c_2^{\Sigma''n}$ & $-0.000167277$ & $0.00180209$ & $-0.000223599$  & $0.0820897$& $0.0321689$& $0.0476438$\\
$c_3^{\Sigma''n}$ & -- & -- & -- & $-0.0148368$& $-0.00881948$& $-0.0170447$\\
$c_4^{\Sigma''n}$ & -- & -- & -- & $0.000990728$& $0.000540511$& $0.00152533$\\
$c_5^{\Sigma''n}$ & -- & -- & -- & $-1.50839\times10^{-8}$ &$-3.05396\times10^{-8}$ & $-1.37901\times10^{-7}$\\\botrule
\end{tabular}
\renewcommand{\arraystretch}{1.0}
\caption{Fit coefficients for the nuclear
response functions $\F_{p,n}^{\Sigma'_L}$ and $\F_{p,n}^{\Sigma''_L}$ for the relevant isotopes of fluorine, sodium, and xenon.  In analogy to Table~\ref{tab:fits}, the fit
functions are $\F(u) = e^{-\frac{u}{2}}\sum_{i} c_i u^i$, with nonzero coefficients as indicated. The results for the other isotopes considered in this work are listed in Tables~\ref{tab:fits_SD_2}--\ref{tab:fits_SD_4}.}
\label{tab:fits_SD_1}
\end{table*}

\begin{table*}[t]
\centering
\renewcommand{\arraystretch}{1.3}
\begin{tabular}{lcccc}
\toprule
Isotope	& \multicolumn{4}{c}{$^{133}$Cs} \\
$L$ & $1$ & $3$ &$5$ & $7$ \\
\colrule
$c_0^{\Sigma'p}$ & $-0.253012$ & -- & --  & -- \\
$c_1^{\Sigma'p}$ & $0.483027$ & $0.104388$ & --  & -- \\
$c_2^{\Sigma'p}$ & $-0.164531$ & $-0.08238$ & $-0.0150628$  & -- \\
$c_3^{\Sigma'p}$ & $0.0168134$ & $0.0118925$ & $0.00856552$  & $0.000657954$ \\
$c_4^{\Sigma'p}$ & $-0.00048879$ & $-0.000423071$ & $-0.000519134$ & $-0.000651735$ \\
$c_5^{\Sigma'p}$ & $-5.62349\times10^{-8}$ & $-3.52071\times10^{-8}$ & $2.07474\times10^{-8}$ & $4.02019\times10^{-8}$ \\\colrule
$c_0^{\Sigma'n}$ & $0.00070445$ & -- & --  & -- \\
$c_1^{\Sigma'n}$ & $-0.00520619$& $-0.00507773$ & --  & -- \\
$c_2^{\Sigma'n}$ & $0.00351738$& $0.00295876$ & $0.000728257$  & -- \\
$c_3^{\Sigma'n}$ & $-0.00069372$ & $-0.000444073$ & $-0.000228224$ & $-0.0000513882$ \\
$c_4^{\Sigma'n}$ & $0.000060668$ & $0.0000235555$ & $0.000018572$ & $6.60564\times10^{-6}$ \\
$c_5^{\Sigma'n}$ & $-1.0888\times10^{-6}$ & $-9.09827\times10^{-7}$ & $-4.68954\times10^{-7}$  & $-2.43844\times10^{-7}$ \\\colrule
$c_0^{\Sigma''p}$ & $-0.178908$ & -- & --  & -- \\
$c_1^{\Sigma''p}$ & $0.0320074$& $0.0904055$ & --  & -- \\
$c_2^{\Sigma''p}$ & $0.0211378$& $0.0034629$& $-0.0137503$ & -- \\
$c_3^{\Sigma''p}$ &  $-0.00419937$& $-0.00308878$ & $-0.00344057$ & $0.000615352$ \\
$c_4^{\Sigma''p}$ &  $0.000173592$& $0.000141839$ & $0.000290862$ & $0.000607814$\\
$c_5^{\Sigma''p}$ & $-6.77831\times10^{-8}$ & $-2.95736\times10^{-8}$ & $1.61033\times10^{-8}$  & $1.93527\times10^{-8}$ \\\colrule
$c_0^{\Sigma''n}$ & $0.000498115$& -- & --  & -- \\
$c_1^{\Sigma''n}$ &$-0.000408223$ & $-0.00439751$ & --  & -- \\
$c_2^{\Sigma''n}$ & $-0.000741592$&$0.00230722$ & $0.000664811$  & -- \\
$c_3^{\Sigma''n}$ & $0.000215744$ & $-0.000355182$ & $-0.000138555$ & $-0.0000480682$ \\
$c_4^{\Sigma''n}$ &  $-0.0000124709$& $0.0000227478$ & $8.67291\times10^{-6}$ & $6.82111\times10^{-6}$ \\
$c_5^{\Sigma''n}$ &  $-1.28214\times10^{-7}$& $-2.62241\times10^{-7}$ & $-2.143\times10^{-7}$  & $-1.52006\times10^{-7}$ \\\botrule
\end{tabular}
\renewcommand{\arraystretch}{1.0}
\caption{Same as Table~\ref{tab:fits_SD_1}, for cesium.}
\label{tab:fits_SD_2}
\end{table*}

\begin{table*}[t]
\centering
\renewcommand{\arraystretch}{1.3}
\begin{tabular}{lccc}
\toprule
Isotope	 &\multicolumn{3}{c}{$^{127}$I}\\
$L$ & $1$ &$3$ &$5$\\
\colrule
$c_0^{\Sigma'p}$ & $0.231258$ & -- & --\\
$c_1^{\Sigma'p}$ & $-0.374391$ & $-0.153173$ & --\\
$c_2^{\Sigma'p}$ & $0.195962$ & $0.105378$ & $0.0743581$\\
$c_3^{\Sigma'p}$ & $-0.0342014$ & $-0.0228849$ & $-0.0234546$\\
$c_4^{\Sigma'p}$ & $0.00162438$ & $0.00130854$ & $0.00188104$\\
$c_5^{\Sigma'p}$ & $-3.37595\times10^{-7}$ & $-2.56507\times10^{-8}$ & $-9.24252\times10^{-8}$\\\colrule
$c_0^{\Sigma'n}$ & $0.0205005$ & -- & --\\
$c_1^{\Sigma'n}$ & $-0.0362175$ & $-0.00369561$ & --\\
$c_2^{\Sigma'n}$ & $0.0174239$ & $0.00235829$ & $0.0000803278$\\
$c_3^{\Sigma'n}$ & $-0.00285902$ & $-0.000383903$ & $-0.0000110023$\\
$c_4^{\Sigma'n}$ & $0.000174649$ & $0.0000204291$ & $2.57961\times10^{-8}$\\
$c_5^{\Sigma'n}$ & $-2.13335\times10^{-6}$ & $-6.28715\times 10^{-7}$ & $-4.55316\times 10^{-8}$\\\colrule
$c_0^{\Sigma''p}$ & $0.163523$ & -- & --\\
$c_1^{\Sigma''p}$ & $-0.125749$ & $-0.132651$ & --\\
$c_2^{\Sigma''p}$ & $0.0450115$ & $0.0668207$ & $0.0678788$\\
$c_3^{\Sigma''p}$ & $-0.00624361$ & $-0.0124938$ & $-0.0207994$\\
$c_4^{\Sigma''p}$ & $0.000245811$ & $0.000614695$ & $0.00171886$\\
$c_5^{\Sigma''p}$ & $-1.85038\times10^{-7}$ & $-1.48556\times 10^{-8}$ & $-3.6698\times10^{-8}$\\\colrule
$c_0^{\Sigma''n}$ & $0.0144959$ & -- & --\\
$c_1^{\Sigma''n}$ & $-0.017996$ & $-0.00320049$ & --\\
$c_2^{\Sigma''n}$ & $0.00679698$ & $0.00161307$ & $0.0000733265$\\
$c_3^{\Sigma''n}$ & $-0.000985306$ & $-0.000240205$ & $-0.0000296856$\\
$c_4^{\Sigma''n}$ & $0.0000461489$  & $0.0000177136$ & $2.66146\times 10^{-6}$\\
$c_5^{\Sigma''n}$ & $-2.6458\times 10^{-7}$ & $-1.81495\times 10^{-7}$ & $-2.07467\times10^{-8}$\\\botrule
\end{tabular}
\renewcommand{\arraystretch}{1.0}
\caption{Same as Table~\ref{tab:fits_SD_1}, for iodine.}
\label{tab:fits_SD_3}
\end{table*}

\begin{table*}[t]
\centering
\renewcommand{\arraystretch}{1.3}
\begin{tabular}{lccccc}
\toprule
Isotope	& \multicolumn{5}{c}{$^{73}$Ge}\\
$L$ & $1$ & $3$ &$5$ & $7$ & $9$\\
\colrule
$c_0^{\Sigma'p}$ & $0.0257064$ & -- & --  & -- & --\\
$c_1^{\Sigma'p}$ & $-0.0418759$& $-0.00920991$ & --  & -- & --\\
$c_2^{\Sigma'p}$ & $0.015169$& $0.00417235$& $0.000347229$  & -- & --\\
$c_3^{\Sigma'p}$ &  $-0.00163883$& $-0.000562315$ & $-0.0000416612$ & $-0.0000106207$ & --\\
$c_4^{\Sigma'p}$ & $0.000045204$ & $0.0000198897$ & $3.28572\times10^{-6}$ & $7.80947\times10^{-7}$ &$0.0000282336$\\\colrule
$c_0^{\Sigma'n}$ & $0.353305$ & -- & --  & -- & --\\
$c_1^{\Sigma'n}$ & $-0.562061$& $-0.233908$ & --  & -- & --\\
$c_2^{\Sigma'n}$ & $0.181791$& $0.117078$& $0.0549077$ & -- & --\\
$c_3^{\Sigma'n}$ & $-0.0180905$ & $-0.0142114$ & $-0.0112771$ & $-0.00718702$ & --\\
$c_4^{\Sigma'n}$ & $0.00051239$ & $0.000451122$ & $0.000469159$ & $0.000528465$ &$0.000752534$\\\colrule
$c_0^{\Sigma''p}$ & $0.0181769$ & -- & --  & -- & --\\
$c_1^{\Sigma''p}$ & $-0.0174535$& $-0.00797608$ & --  & -- & --\\
$c_2^{\Sigma''p}$ & $0.00405522$& $0.00361834$& $0.000316977$ & -- & --\\
$c_3^{\Sigma''p}$ & $-0.00028759$ & $-0.000359967$ & $-0.0000879456$ & $-9.93462\times10^{-6}$ & --\\
$c_4^{\Sigma''p}$ & $6.37115\times10^{-6}$ & $6.89172\times10^{-6}$ & $1.79986\times10^{-6}$ & $5.84383\times10^{-7}$&$0.0000267848$\\\colrule
$c_0^{\Sigma''n}$ & $0.249824$ & -- & --  & -- & --\\
$c_1^{\Sigma''n}$ & $-0.205762$& $-0.202568$ & --  & -- & --\\
$c_2^{\Sigma''n}$ & $0.045436$& $0.0675417$ & $0.0501235$ & -- & --\\
$c_3^{\Sigma''n}$ & $-0.00335668$ & $-0.00613658$ & $-0.00769287$ & $-0.00672282$ & --\\
$c_4^{\Sigma''n}$ & $0.0000723721$ & $0.00015611$ & $0.000256944$ & $0.000395465$& $0.000713918$\\\botrule
\end{tabular}
\renewcommand{\arraystretch}{1.0}
\caption{Same as Table~\ref{tab:fits_SD_1}, for germanium.}
\label{tab:fits_SD_4}
\end{table*}
 
\end{widetext}

\end{document}